\documentclass[11pt,a4paper]{article}
\usepackage{t1enc}
\usepackage[utf8]{inputenc}
\usepackage[english]{babel}
\normalfont
\usepackage{amsmath}
\usepackage{amssymb}
\usepackage{amsthm}

\usepackage{mathabx}
\usepackage{bbm}
\usepackage{mathrsfs}
\usepackage{pifont}
\usepackage{hyperref}
\usepackage{pgf}
\usepackage{graphicx}
\usepackage[inline]{enumitem}
\usepackage{tikz-cd}
\setlist{nosep, itemsep=.1cm, topsep=.1cm}

\newcommand{\be}[0]{\begin{equation}}
\newcommand{\ee}[0]{\end{equation}}
\newcommand{\btkz}[0]{\begin{tikzcd}}
\newcommand{\etkz}[0]{\end{tikzcd}}

\newcommand*{\textcal}[1]{%
  \textit{\large \fontfamily{pzc}\selectfont#1}%
}

\numberwithin{equation}{section}

\theoremstyle{plain}
\newtheorem{theorem}{Theorem}[section]

\theoremstyle{definition}

\begin{document}

\vspace*{-1cm}
\thispagestyle{empty}
\vspace*{1.5cm}

\begin{center}
{\Large 
{\bf The anomaly field theories of six-dimensional (2,0) superconformal theories }}
\vspace{2.0cm}

{\large Samuel Monnier}
\vspace*{0.5cm}

Section de Mathématiques, Université de Genève\\
2-4 rue du Lièvre, 1211 Genève 4, Switzerland\\
samuel.monnier@gmail.com

\vspace*{1cm}

{\bf Abstract}
\end{center}

We construct 7-dimensional quantum field theories encoding the anomalies of conformal field theories with (2,0) supersymmetry in six dimensions. We explain how the conformal blocks of the (2,0) theories arise in this context. A result of independent interest is a detailed specification of the data required to define a (2,0) theory with topologically non-trivial spacetime and R-symmetry bundle.

\newpage

\tableofcontents

\section{Introduction and summary}

Conformal field theories with $(2,0)$ supersymmetry in dimension six (henceforth $(2,0)$ SCFTs, see for instance \cite{Moore2012}), play a central role in our understanding of the non-perturbative physics of lower-dimensional supersymmetric quantum field theories. They are however notoriously difficult to study because of their intrinsically quantum nature: they do not admit a semiclassical limit in which perturbative methods would apply. At the price of breaking the conformal symmetry, it is possible to reduce them to essentially free theories in the IR, by turning on generic Coulomb branch parameters. While this process drastically changes the properties of theory, 't Hooft anomaly matching shows that the gravitational and R-symmetry anomalies are invariant. These anomalies are therefore computable quantities offering a window into the strongly coupled regime of $(2,0)$ SCFTs. 

The local gauge and R-symmetry anomalies were computed in \cite{Harvey:1998bx, Intriligator:2000eq}, while their global counterparts have been recently derived in \cite{Monnier:2014txa}. There is however much more information in the anomalies of $(2,0)$ SCFTs than was extracted by these papers. For instance, a $(2,0)$ SCFT on a 6-manifold $M$ does not generally have a single partition function, but rather a vector of "conformal blocks", of dimension $d = |H^3(M;\Gamma)|^{1/2}$, where $\Gamma$ is the finite group obtained as the quotient of the weight lattice by the root lattice of the ADE Lie algebra defining the SCFT. Under transformations disconnected from the identity, the vector of conformal blocks is transformed by a $U(d)$ element. \cite{Harvey:1998bx, Intriligator:2000eq} considered only infinitesimal transformations, and \cite{Monnier:2014txa} only transformations leaving the vector of conformal blocks invariant up to a phase. Moreover, there are Hamiltonian anomalies, affecting the state space of the theory on a 5-dimensional manifold, as well as more exotic anomalies affecting the objects the theory associates to lower dimensional manifolds. 

A prior, it is a challenge to describe all the anomalies and the consistency relations they obey. A recent insight addressing this problem is the notion of anomaly field theory \cite{Freed:2014iua, Moore, Moore2012, 2014arXiv1406.7278F, 2014arXiv1409.5723F, Monnierd}: all the anomalies of a $d$-dimensional quantum field theory are encoded in an extended field theory in dimension $d+1$ (or more precisely in an equivalence class thereof), the \emph{anomaly field theory}. Moreover, the consistency constraints that anomalies satisfy are nothing but the requirement that the anomaly field theory is a field theory functor, in the Atiyah-Segal sense. This formalism also naturally includes anomalous "relative quantum field theories" \cite{Freed:2012bs} which do not have a unique partition function or state space, such as the chiral conformal field theories in two dimensions or the $(2,0)$ SCFTs to be discussed here. 

The main result of the present paper is the construction of anomaly field theories for the $(2,0)$ SCFT, as non-extended quantum field theories. The 7-dimensional anomaly quantum field theories are the product of certain invertible field theories with a discretely gauged Wu Chern-Simons theory, constructed in \cite{Monnier:2016jlo}. While the invertible field theories can easily be formulated as extended field theories (see for instance \cite{Freed:1994ad, 2014arXiv1406.7278F}), the Wu Chern-Simons theory is currently known only as an ordinary field theory. The anomaly field theories, in their current non-extended formulation, therefore only contain information about the anomalies of the conformal blocks of the $(2,0)$ SCFTs. It would be very interesting to extend them at least to codimension 2, to study Hamiltonian anomalies, but this is beyond the scope of the present work.

All the quantum field theories to be discussed here are Euclidean. A suitable Wick rotation relates the correlation functions of the Lorentzian and Euclidean theories, and therefore their anomalies as well. The focus of this paper is on gauge and gravitational anomalies, but we will comment on conformal anomalies at the end of this introduction. We will now recall the notion of anomaly field theory and summarize the results of the paper in more detail. A more elaborate discussion of the concept of anomaly field theory can be found in \cite{Monnierd}.

\paragraph{Anomaly field theories}  The thesis underlying the concept of anomaly field theory is that a $d$-dimensional anomalous field theory is nothing but a "field theory taking value in a certain $d+1$ field theory", the \emph{anomaly field theory}.

To understand what this means, recall that a $d$-dimensional quantum field theory assigns in particular a complex number, the \emph{partition function}, to any closed $d$-dimensional manifold $M^d$, and a Hilbert space, the \emph{state space}, to any closed $d-1$-dimensional manifold $M^{d-1}$. A "$d$-dimensional field theory $\mathcal{F}$ taking value in a $d+1$-dimensional field theory $\mathcal{A}$" assigns an element of the Hilbert space $\mathcal{A}(M^d)$ to $M^d$. Its partition function is therefore a vector rather than a complex number. Similarly, its state space $\mathcal{F}(M^{d-1})$ is not a Hilbert space, but rather an object in the category assigned by $\mathcal{A}$ to $M^{d-1}$ (which can be physically pictured as the category of boundary conditions of $\mathcal{A}$). Of course, these assignments are subject to consistency conditions. Those can be formalized neatly by seeing $\mathcal{A}$ as a functor from a (higher) bordism category to the (higher) category of Hilbert spaces. $\mathcal{F}$ is then a natural transformation from $\mathcal{A}$ to the trivial $d+1$-dimensional field theory functor. We refer the reader to \cite{Monnierd} for an explanation of these claims.

Familiar anomalous quantum field theories, such as chiral fermions, have invertible field theories. Recall that a $d+1$-dimensional quantum field theory $\mathcal{A}$ is called \emph{invertible} when the objects it assigns to $d+1$- and $d$-dimensional manifolds of various dimensions are all invertible. For instance, its partition function on a closed $d+1$-dimensional manifold should be a non-vanishing complex number, and its state space on a closed $d$-dimensional manifold should be a 1-dimensional Hilbert space, i.e. a Hermitian line, which is invertible with respect to the tensor product operation. Anomalous field theories with invertible anomaly field theories have therefore partition functions taking value in a Hermitian line. As Hermitian lines can be non-canonically be identified with $\mathbb{C}$, their partition functions can be identified with complex numbers at the price of unnatural choices. If a unitary symmetry is present, it acts on $\mathcal{A}(M^d)$ by multiplication by a phase. The partition function of $\mathcal{F}$, being a vector in $\mathcal{A}(M^d)$, gets multiplied by this phase and therefore fails to be invariant under the symmetry. This shows how the conventional picture of anomalies as symmetry breaking phenomena is recovered in this formalism.

Examples of anomalous quantum field theories with value in non-invertible anomaly field theories are provided by rational chiral conformal theories. They generally do not admit a single partition function, but rather a vector of "conformal blocks". This vector of conformal blocks takes value in the state space of a Reshetikhin-Turaev topological field theory, constructed out of the modular tensor category of representations of the relevant chiral vertex algebra. In the particular case of chiral WZW models, the Reshetikhin-Turaev theory is quantum Chern-Simons theory, and the observation above dates back to \cite{Witten:1988hf}.

\paragraph{The anomaly field theories of $(2,0)$ SCFTs} The $(2,0)$ SCFTs in dimension six studied in the present paper are similar to chiral conformal field theories in the sense that they generally have a vector of conformal blocks rather than a single partition function. Accordingly, their 7-dimensional anomaly field theories are non-invertible. (The only exception is the $E_8$ theory.)  

The anomaly field theory to be described is defined on 7-dimensional manifolds carrying all the data necessary to define a $(2,0)$ SCFT in dimension six. We refer to these manifolds as $(2,0)$-manifolds in the following, see Section \ref{SecPrelim} for definitions. $(2,0)$-manifolds are in particular endowed with a rank 5 R-symmetry bundle $\mathscr{N}$. For the $(2,0)$ SCFT based on the Lie algebra $\mathfrak{g}$, we find that the anomaly field theory is
\be
\label{EqAFTGenSCFTIntro}
\textcal{An}_{\mathfrak{g}} = \left(\textcal{DF}^{\frac{1}{2}}_{f} \right)^{\otimes (-r_\mathfrak{g})} \otimes \left( \textcal{DF}^{\frac{1}{4}}_\sigma \right)^{\otimes (-r_\mathfrak{g})} \otimes \textcal{An}_{\rm HWZ} \otimes  \overline{\textcal{WCS}}_{\rm G}[\Lambda_\mathfrak{g}, 0] \;,
\ee
with
\be
\label{EqAFTHWZGenIntro}
\textcal{An}_{\rm HWZ} =  \left( \textcal{WCS}_{\rm P}[\mathbb{Z}, -2\check{b}]\right)^{\otimes \frac{r_\mathfrak{g} {\rm h}_\mathfrak{g}}{2}} \otimes
\left(\textcal{BF}[-2\check{b}, \check{C}']\right)^{\otimes \frac{r_\mathfrak{g} {\rm h}_\mathfrak{g}}{2}} \otimes 
\left( \textcal{CSp}_2[\check{b}] \right)^{\otimes \frac{|\mathfrak{g}| {\rm h}_\mathfrak{g}}{6}} \;.
\ee
The notation is as follows. Each factor corresponds to a quantum field theory, and the tensor product operation corresponds physically to taking non-interacting copies of the relevant field theories on the same spacetime. $r_\mathfrak{g}$, $\Lambda_\mathfrak{g}$, ${\rm h}_\mathfrak{g}$ and $|\mathfrak{g}|$ denote respectively the rank, root lattice, dual Coxeter number and dimension of $\mathfrak{g}$. 

$\textcal{DF}^{\frac{1}{2}}$ is a "half Dai-Freed theory" \cite{Dai:1994kq}, a 7-dimensional invertible field theory describing the anomalies of 6-dimensional symplectic Majorana-Weyl fermions valued in the spinor bundle of $TM \otimes \mathscr{N}$, where $M$ is the spacetime. As the 6d SCFT contains $r_\mathfrak{g}$ tensor multiplets on the Coulomb branch, each involving one such fermion with negative chirality, we have $r_\mathfrak{g}$ non-interacting copies of the complex conjugate of $\textcal{DF}^{\frac{1}{2}}$, as denoted by the tensor product with negative exponent. We discuss this theory in more detail in Section \ref{SecAFTM5WV}. The second factor $\textcal{DF}^{\frac{1}{4}}$ is a "quarter Dai-Freed theory" associated to the signature Dirac operator. It essentially describes the anomaly of the self-dual fields present in the tensor multiplets. This field theory is discussed in Section \ref{SecAFTM5WV} as well.

$\overline{\textcal{WCS}}_{\rm G}[\Lambda_\mathfrak{g}, 0]$ is a discretely gauged Wu Chern-Simons theory, constructed in \cite{Monnier:2016jlo} and discussed in more details in Section \ref{SecDiscGaugWCS}. $\overline{\textcal{WCS}}_{\rm G}[\Lambda_\mathfrak{g}, 0]$ is the only non-invertible factor in \eqref{EqAFTGenSCFTIntro}. Its state space on a 6-manifold $M$ has dimension $|H^3(M;\Lambda^\ast_\mathfrak{g}/\Lambda_\mathfrak{g})|^{1/2}$, which is an integer because of the perfect skew-symmetric pairing on $H^3(M;\Lambda^\ast_\mathfrak{g}/\Lambda_\mathfrak{g})$. This is consistent with expected dimension of the vector of conformal blocks of the $(2,0)$ SCFT \cite{Witten:1998wy}. In the presence of torsion in $H^3(M;\mathbb{Z})$, the Heisenberg module structure on the space of conformal blocks is however different than what was conjectured in \cite{Witten:1998wy}, see the discussion in Section \ref{SecConfBlocks}. 

Finally, $\textcal{An}_{\rm HWZ}$ is the anomaly field theory associated to the "Hopf-Wess-Zumino terms" \cite{Intriligator:2000eq} present on the Coulomb branch of the $(2,0)$ SCFT. It is a product of three distinct invertible quantum field theories, as detailed in \eqref{EqAFTHWZGenIntro}, and involves two background fields $\check{b}$ and $\check{C}'$. For simple enough topologies of the $(2,0)$ theory's spacetime and R-symmetry bundle, $\check{b}$ and $\check{C}'$ vanish and $\textcal{An}_{\rm HWZ}$ is trivial.

We first describe the background fields, and then the factors of \eqref{EqAFTHWZGenIntro}. $\check{b}$ is a degree 3 background gauge field whose field strength has half-integral fluxes congruent mod 1 to half the periods of $w_4(\mathscr{N})$, the fourth Stiefel-Whitney class of the R-symmetry bundle $\mathscr{N}$. Recall that the $A_n$ $(2,0)$ SCFTs can be realized as stacks of M5-branes after decoupling the center of mass of the stack. $\mathscr{N}$ is then interpreted as the normal bundle of the stack. When $\mathscr{N}$ is non-trivial, there is no unique way of decoupling the center of mass, and $\check{b}$ encodes a choice of decoupling. When the 4th Stiefel-Whitney class $w_4(\mathscr{N})$ vanishes, we can choose $\check{b} = 0$. Else, it has to be understood as part of the definition of the $(2,0)$ SCFT (see Section \ref{SecDataDefSCFT}). 

On the 6-dimensional spacetime of the $(2,0)$ SCFT, $\check{C}'$ is a background degree 3 abelian gauge field, of which torsion fluxes may have to be turned on to avoid gauge anomalies of the self-dual fields in the $(2,0)$ theory \cite{Witten:1999vg}. In order to describe the most general anomalies of the $(2,0)$ SCFT, $\check{C}'$ should be allowed to be an arbitrary background abelian gauge field on the 7-dimensional spacetime of the anomaly field theory.

We now describe the factors in \eqref{EqAFTHWZGenIntro}. $\textcal{WCS}_{\rm P}[\mathbb{Z}, -2\check{b}]$ is a prequantum Wu Chern-Simons theory \cite{Monnier:2016jlo} based on the lattice $\mathbb{Z}$, with background abelian gauge field $-2\check{b}$. (See Section \ref{SecPreqWCS} for an explanation of the term "prequantum".) The second factor $\textcal{BF}[-2\check{b}, \check{C}']$ is a 7-dimensional prequantum BF theory constructed from the gauge fields $-2\check{b}$ and $\check{C}'$. $\textcal{CSp}_2[\check{b}]$, a "prequantum Chern-Simons-$p_2$" theory, is a new invertible 7-dimensional field theory. It is essentially a quadratic Chern-Simons theory with degree 3 abelian gauge field $\check{b}$. However, because of the shift in the quantization of the fluxes of $\check{b}$, such a theory would be ill-defined. The action of the Chern-Simons-$p_2$ theory contains a second term derived from $\frac{1}{4}p_2$, where $p_2$ is the second Pontryagin class. Because of the fractional factor, the action associated to this second term is ill-defined as well. However, they yield together a well-defined action and prequantum theory. The Hopf-Wess-Zumino anomaly field theory and its three components are described in more details in Section \ref{SecAFTHWZ}.

\paragraph{Derivation of the anomaly field theory} The anomaly field theory \eqref{EqAFTGenSCFTIntro} is designed to reproduce the global anomaly of the $(2,0)$ SCFT computed in \cite{Monnier:2014txa}. The global anomaly of \cite{Monnier:2014txa} determines the partition function of the anomaly field theory, and we find a natural way of consistently completing this data to a quantum field theory. Elementary properties of field theory functors imply that this completion is essentially unique (see the discussion below).

There are however two shortcomings in the derivation. First, the global anomaly was derived in \cite{Monnier:2014txa} for the $A$ series, using the realization of the $A_n$ $(2,0)$ SCFT on a stack of M5-branes, but only conjectured for the $D$ and $E$ series. The same is restrictions apply in the present paper.

A second shortcoming is the following. The derivation of the anomaly field theory to be presented below is valid only if every 7-dimensional $(2,0)$-manifold $M$ is the boundary of a 8-dimensional $(2,0)$-manifold. In Appendix \ref{SecCobGroup20Struct} and Section \ref{SecCat20Man}, we show that this is true when
\be
\label{EqCondTopRSymBun}
w_2(TM) w_3(TM) = 0 \;,
\ee
where $w_i$ are the Stiefel-Whitney classes. We do not know whether there exists a 7-dimensional $(2,0)$-manifold $\tilde{M}$ that does not bound an 8-dimensional $(2,0)$-manifold, but should it exist, it would necessarily violate \eqref{EqCondTopRSymBun}. In this case, it may be that the correct anomaly field theory differs from the one presented here by a sign on $\tilde{M}$. We therefore restrict our discussion to $(2,0)$ SCFTs whose R-symmetry bundle satisfy \eqref{EqCondTopRSymBun}, and to anomalies that can be computed using 7-dimensional $(2,0)$-manifolds satisfying \eqref{EqCondTopRSymBun}. 

\paragraph{Conformal anomaly} We should emphasize that the anomaly field theory described in the present paper describes the gravitational and R-symmetry anomalies of the $(2,0)$ SCFTs, but \emph{not} directly their conformal anomalies. The main tool for deriving the anomaly field theory is anomaly inflow from M-theory onto a stack of M5-branes. M-theory is not conformally invariant; conformal invariance is obtained only after a decoupling limit. As a result, there is no reason to expect the construction of the present paper to capture directly the conformal anomaly. 

It is expected that supersymmetry should relate the conformal anomaly to the local gravitational and R-symmetry anomalies, which are computable from the anomaly field theory. This relation is however still elusive.\footnote{We thank Ken Intriligator for pointing this out to us.} The conformal anomalies of $(2,0)$ SCFTs have been computed from first principles recently in \cite{Cordova:2015vwa}. \\

The paper is organized as follows. In Section \ref{SecPrelim}, we spell out in detail the data required to define a $(2,0)$ SCFT on a 6-dimensional manifold, yielding the notion of $(2,0)$-manifold. For the purpose of computing anomalies, we also need to consider $(2,0)$-manifolds of dimension 7 and 8. We define morphisms of $(2,0)$-manifolds and the associated category. In Section \ref{SecDiscGaugWCS}, we recall some of the results of \cite{Monnier:2016jlo} about Wu Chern-Simons theory. We find a relation between Wu Chern-Simons theories whose gauge groups are related by lattice decompositions, which is crucial to perform the subtraction of the center of mass anomaly. In Section \ref{SecAFTStack}, we describe the anomaly field theory of a stack of M5-branes, decomposing it into the product of an anomaly field theory due to the worldvolume of the M5-branes and an anomaly field theory due to the Hopf-Wess-Zumino terms of \cite{Intriligator:2000eq}. In Section \ref{SecAFTCM}, we describe the anomaly field theory of the center of mass tensor multiplet. The anomaly field theory of the $A_n$ SCFT is then derived in Section \ref{Sec7dimform}. We use it to conjecture the anomaly field theories of SCFTs in the $D$ and $E$ series. This section also contains a brief discussion of the relation between the defects of the $(2,0)$ SCFT and the defects of its anomaly field theory. In Section \ref{SecConfBlocks}, we discuss the implication of our results for the conformal blocks of the $(2,0)$ SCFT. Appendix \ref{AppDiffCohom} reviews the differential cohomology model of abelian gauge fields. Wu structures and Euler structures are presented in Appendix \ref{AppWuStruct} and \ref{AppEulerStruct}, respectively. Appendix \ref{SecCobGroup20Struct} contains a proof that the cobordism group of 7-dimensional $(2,0)$-manifolds subject to \eqref{EqCondTopRSymBun} vanishes.

\section{(2,0)-manifolds}

\label{SecPrelim}

We assume that the reader is familiar with the differential cohomology model of (higher) abelian gauge fields, briefly reviewed in Appendix \ref{AppDiffCohom}. In this model, degree $p-1$ abelian gauge fields (with degree $p$ field strengths) are degree $p$ differential cocycles. The gauge equivalence classes of degree $p-1$ abelian gauge fields are then in bijection with degree $p$ differential cohomology classes. Shifted differential cocycles model gauge fields whose field strength may have fractional fluxes, such as the M-theory C-field. Differential cocycles will always be written with a caron ($\check{C}$).

\subsection{Data required to define a (2,0) SCFT and its anomaly field theory}

\label{SecDataDefSCFT}

We start by recalling the topological and geometrical data required for the definition of a (Euclidean) (2,0) SCFT on a manifold $M$. The same data is required for the definition of the corresponding 7-dimensional anomaly field theory. We will in fact need to consider manifolds endowed with such data in dimensions up to 8. 
The data required to define a (2,0) SCFT is composed of the following:
\begin{enumerate}
  \item A choice of a Lie algebra $\mathfrak{g}$ of A, D or E-type. This fixes the gauge symmetry of the theory.
	\item An orientation, a smooth structure, a Riemannian metric on the manifold $M$, which we will take to be compact for simplicity.
	\item A rank 5 bundle $\mathscr{N}$ over $M$ endowed with an inner product and a compatible connection, satisfying
\be
\label{EqRelSWClassesTMN}
w_1(TM) = w_1(\mathscr{N}) = 0 \;, \quad w_2(TM) + w_2(\mathscr{N}) = 0 \;, \quad w_5(\mathscr{N}) = 0 \;.
\ee 
The first equalities ensure that both $M$ and $\mathscr{N}$ are orientable. The second equality implies that $TM \oplus \mathscr{N}$ is spin. $w_5$ is the reduction mod 2 of the Euler class $e(\mathscr{N})$, which is $\mathbb{Z}_2$-torsion, so the last equality is equivalent to $e(\mathscr{N}) = 0$. It is a consequence of the first two equalities in dimensions 7 or lower, as explained in Appendix A of \cite{Monnierb}. We write $\pi: \mathscr{N} \rightarrow M$ for the bundle projection.
From the point of view of the (2,0) theory, $\mathscr{N}$ is the R-symmetry bundle in which the Coulomb branch parameters take value. In the case of the M5-brane realization of the $A_n$ theory, $\mathscr{N}$ is the normal bundle of the stack of M5-branes.

	\item A spin structure on $TM \oplus \mathscr{N}$. This spin structure is necessary to define the fermionic fields in the free tensor multiplets appearing on the Coulomb branch. Note that we do \emph{not} need $M$ to be spin.

	\item An Euler structure on $\mathscr{N}$ (see Appendix \ref{AppEulerStruct}). The requirement that $e(\mathscr{N}) = 0$ ensures that Euler structures on $\mathscr{N}$ exist \cite{Diaconescu:2003bm, Monnierb}. Concretely, an Euler structure provides an integral cocycle $a$ representing the top cohomology class of the fibers of $\tilde{M}$, the 4-sphere bundle over $M$ associated to $\mathscr{N}$. 
	
We also need a differential cocycle refinement $\check{a}$ of $a$, i.e. a differential cocycle $\check{a}$ whose characteristic is $a$. We will take it to be of the form 
\be
\label{EqChoicFormGlobAngCoc}
\check{a} = \frac{1}{2} \check{e}(T_V \tilde{M}) + \pi^\ast(\check{a}') \;.
\ee
$\check{e}(T_V \tilde{M})$ is the differential cocycle associated to the Euler class of the vertical tangent bundle $T_V \tilde{M}$ and the connection on $T_V \tilde{M}$ inherited from $\mathscr{N}$. (See Theorem 2.2 of \cite{springerlink:10.1007/BFb0075216} for more detail about how to associate a differential cocycle to a bundle with connection and a characteristic class.) $\check{a}'$ is a differential cocycle on $M$ with harmonic curvature. The harmonicity condition uniquely fixes the curvature of $\check{a}$.  As the Euler class may not be divisible by 2, $\check{a}'$ may be a shifted (and therefore non-vanishing) differential cocycle in order to ensure that $\check{a}$ is unshifted. Like $a$, $\check{a}$ integrates to $1$ on the 4-sphere fibers of $\tilde{M}$. 

The Euler structure should be thought of as a way of decomposing degree 4 cohomology classes on $\tilde{M}$ into "fiberwise" and "longitudinal" components. $\check{a}$ extends this decomposition to degree 4 differential cocycles (representing degree 3 abelian gauge fields). In the M-theory realization of the $A_n$ SCFTs, such a decomposition is necessary in order to decouple the center of mass of the stack of M5-branes \cite{Monnier:2014txa}. We explain below that in favorable cases, like for instance when $\mathscr{N}$ is trivial, the Euler structure and $\check{a}$ can be chosen canonically. 

The results of \cite{Witten:1999vg} imply that
\be
\label{EqDefbIntGlobAngForm}
\check{b} := \frac{1}{2} \pi_\ast(\check{a} \cup \check{a})
\ee
is a degree 4 differential cocycle on $M$ shifted by $w_4(\mathscr{N})$.

	\item A degree 4 differential cocycle $\check{C}_{M}$ shifted by the degree 4 Wu class of $TM$, see Appendix \ref{AppWuStruct}. $\check{C}_{M}$ is a higher abelian gauge field coupling to the self-dual fields in the tensor multiplets present on the Coulomb branch of the (2,0) SCFT. In the M-theory realization of the $A_n$ SCFT $\check{C}_M$ is the effective C-field on the worldvolume of the stack of M5-branes \cite{Monnierb}. Note that unless the dimension of $M$ is 8, the Wu class vanishes, and $\check{C}_M$ is in fact an unshifted differential cocycle. On a 6-manifold supporting a (2,0) SCFT, it would be natural to take $\check{C}_{M}$ to vanish, but an analogue of the Freed-Witten anomaly affecting the self-dual fields may require its characteristic to be a certain 2-torsion class \cite{Witten:1999vg, Monniera}. In turn, this implies that we have to allow for arbitrary C-fields on 7-dimensional manifolds in order to be able to compute all the anomalies in 6 dimensions.
\end{enumerate}
In special cases, the data above can be trimmed down. For instance, assume that $\mathscr{N} \simeq M \times \mathbb{R}^5$ is trivial with the canonical connection. Then $M$ has to be spin. $T_V \tilde{M} \simeq TS^4 \times M$, so the Euler class is divisible by 2 and we can take $\check{a}' = 0$. $\check{a}$ is the pullback of a top differential cocycle on the 4-sphere, whose curvature is fixed by the harmonicity condition. The twisting construction, commonly used in order to obtain supersymmetric gauge field theories from the compactification of a (2,0) SCFT, requires however in general a non-trivial R-symmetry bundle. Then, if at least $w_4(\mathscr{N}) = 0$, $\check{b}$ is unshifted and, on 6- and 7-dimensional manifolds we can choose $\check{a}$ such that $\check{b} = \check{C}_M$. As explained above, in the absence of the Freed-Witten-like anomaly we may also choose $\check{C}_M = 0$.

In addition, we will also choose a Wu structure of degree 4 on $TM$ if ${\rm dim}(M) < 8$. Wu structures should be thought of as generalizations of spin structures and are described in Appendix \ref{AppWuStruct}. In the same way as any oriented manifold of dimension smaller or equal to 3 admits a spin structure, any manifold of dimension smaller or equal to 7 admits a Wu structure of degree 4, so this does not put restrictions on the manifolds we consider. The theory is independent of the choice of the Wu structure, but the latter will be useful in certain constructions below. 

In the following, we refer to the data above, including the Wu structure, as a $(2,0)$-structure, and to manifolds endowed with $(2,0)$-structures as $(2,0)$-manifolds. 

\subsection{The category of $(2,0)$-manifolds}

\label{SecCat20Man}

For two $(2,0)$-manifolds $M$ and $N$ of dimension respectively smaller and strictly smaller than 8, a morphism of $(2,0)$-manifolds from $M$ to $N$ is a smooth orientation preserving isometric embedding compatible with the rest of the $(2,0)$-structures, and similarly for two $(2,0)$-manifolds of dimension 8. There is clearly no morphism from $M$ to $N$ if ${\rm dim}(M) > {\rm dim}(N)$.

When $M$ has dimension strictly smaller than 8 and $N$ has dimension 8, the definition of morphisms is less straightforward. $M$ carries a Wu structure, but $N$ does not, and $\check{C}_N$ is shifted by the Wu class of $N$ while $\check{C}_M$ is unshifted. We define the morphisms between from $M$ to $N$ to be again smooth orientation preserving isometric embeddings compatible with the rest of the $(2,0)$-structure, subject to the following compatibility condition.  

As explained in Appendix \ref{AppWuStruct}, the Wu structure on $M$ can be pictured as a trivialization $\eta$ of the Wu cocycle $\nu_M = w_4(TM) + w_2^2(TM)$ (itself obtained via the pullback of a representing cocycle on the associated classifying space). $N$ also comes with a Wu cocycle $\nu_N$, which however is not necessarily trivial, because the degree 4 Wu class of an 8-manifold may be non-zero. Extending the cochain $\eta$ arbitrarily to $N$, we obtain a cocycle 
\be
\label{EqDefMu8Man}
\mu = \nu_N - d\eta
\ee
vanishing on $M$. We require that $\check{C}_N$ is an extension of $\check{C}_M$ as a differential cocycle shifted by $\frac{1}{2} \mu$. $\check{C}_N$ is, as required, shifted by the Wu class, because $\mu$ differs from $\nu_N$ by an exact cocycle.

Armed with the notion of morphism of $(2,0)$-manifolds, we can now consider $(2,0)$-manifolds $M$ with boundary. We require the embedding of $\partial M$ into $M$ to be a morphism of $(2,0)$-manifolds. In addition, for technical reasons, we require that the Riemannian metric is isometric to a direct product in a neighborhood of the boundary.

We will always implicitly restrict ourselves to $(2,0)$-manifolds satisfying the constraint
\be
\label{EqConst20Mfld}
w_2(TM)w_3(TM) = 0 \;.
\ee
As we necessarily have $w_2(\mathscr{N}) = w_2(TM)$ from \eqref{EqRelSWClassesTMN} and $w_3 = {\rm Sq}^1 w_2$, the condition \eqref{EqConst20Mfld} is equivalent to
\be
\label{EqConst20MfldN}
w_2(\mathscr{N})w_3(\mathscr{N}) = 0 \;.
\ee
In Appendix \ref{SecCobGroup20Struct}, we show that the 7-dimensional cobordism group of manifolds with $(2,0)$-structure satisfying \eqref{EqConst20MfldN} vanishes. This means that any 7-dimensional $(2,0)$-manifold satisfying \eqref{EqConst20Mfld} is the boundary of an 8-dimensional $(2,0)$-manifold on which the $(2,0)$-structure extends. The argument of Appendix \ref{SecCobGroup20Struct} do not exclude that the same is true without the constraint \eqref{EqConst20Mfld}. 

If a $(2,0)$-manifold $M$ is spin, then $w_2(TM) = 0$ and \eqref{EqConst20Mfld} is automatically satisfied. More generally, if it is spin$^c$, then $w_2(TM)$ is the reduction of an integral class and $w_3(TM) = {\rm Sq}^1 w_2(TM)$ vanishes, yielding \eqref{EqConst20Mfld} as well.

\section{Discretely gauged Wu Chern-Simons theories}

\label{SecDiscGaugWCS}

We review in this section certain topological field theories on manifolds with Wu structures constructed in \cite{Monnier:2016jlo}, the so called discretely gauged Wu Chern-Simons theories.

On 3-manifolds, one can define Chern-Simons theories with half-integer level \cite{2005math......4524J,2006math......5239J,Belov:2005ze}. These theories depend on a choice of spin structure on the 3-manifold. This statement has a generalization for higher degree abelian gauge fields. On a $4k+3$-dimensional manifold endowed with a degree $2k+2$ Wu structure, one can define a quadratic Wu Chern-Simons theory with half-integral level involving a degree $2k+1$ abelian gauge field. 

Given the classical Wu Chern-Simons action, one can construct invertible field theories, the prequantum Wu Chern-Simons theories. The theories of interest here are obtained by gauging a discrete symmetry of the prequantum Wu Chern-Simons theories. Alternatively, they can be seen as defined by a path integral over discrete gauge fields, akin to Dijkgraaf-Witten theories.

\subsection{The prequantum theory}

\label{SecPreqWCS}

\paragraph{Gauge group} The abelian gauge group and level of a generic abelian Chern-Simons theory can be elegantly encoded in the data of an even lattice $\Lambda$. The gauge group is then the torus $(\Lambda \otimes_{\mathbb{Z}} \mathbb{R})/\Lambda$. $U(1)$ Chern-Simons theory at level $k$ corresponds to the even lattice $\sqrt{2k}\mathbb{Z}$. Similarly, the gauge group and level of a generic abelian spin Chern-Simons theory can be specified by an integral lattice $\Lambda$. The gauge group of a Wu Chern-Simons theory is analogously specified by an integral lattice $\Lambda$. The $2k+1$-form gauge field of the theory can then be modelled by a differential cocycle $\check{C} = (g, C, G)$ taking value in $\Lambda$ (see Appendix \ref{AppDiffCohom} for definitions).

\paragraph{Wu structure} Let $M$ be a 7-dimensional $(2,0)$-manifold. As explained in Appendix \ref{AppWuStruct}, the data of a (degree 4) Wu structure can be encoded in a trivialization of the Wu cocycle: $d\eta = \nu$, where $\eta$ and $\nu$ are $\mathbb{Z}_2$-valued cocycles. Let us lift $\eta$ to a $\mathbb{Z}$-valued cochain $\eta_{\mathbb{Z}}$ and pick a characteristic element $c \in \Lambda$, i.e. an element such that $(c,b) = (b,b)$ modulo 2 for all $b \in \Lambda$. We define the cochains $\eta_{\Lambda} := \eta_{\mathbb{Z}} \otimes c$ and $\nu_\Lambda = d\eta_{\Lambda}$. They can be gathered into a trivial $\Lambda$-valued differential cocycle $\check{\nu} = (\nu_\Lambda, -\eta_\Lambda,0)$.

\paragraph{Lagrangian} The Lagrangian of the theory is the real-valued cocycle
\be
\label{EqLagWuCSTheory}
L(\Lambda, \check{C}) = \frac{1}{2}\left[ \check{C} \cup (\check{C} + \check{\nu}) \right]_{\rm co}
\ee
where $[...]_{\rm co}$ denotes the connexion part of the differential cocycle in the bracket, see Appendix \ref{AppDiffCohom}. $\cup$ is the cup product of differential cocycles, also defined in Appendix \ref{AppDiffCohom}. 

The familiar $U(1)$ Chern-Simons action at level $k$ is recovered when $\Lambda = \sqrt{2k} \mathbb{Z}$ and the gauge field $\check{C} = (g, C, G)$ is topologically trivial ($g = 0$). Then the second term in \eqref{EqLagWuCSTheory} does not contribute modulo integers. Using the cocycle condition $G = dC$, we have
\be
\frac{1}{2} [\check{C} \cup \check{C}]_{\rm co} = \frac{1}{2} C \cup G + \frac{1}{2} H^\wedge_\cup(G, G) \sim \frac{1}{2} C \wedge_\Lambda dC = k \, C \wedge_\mathbb{Z} dC \;,
\ee
where $\sim$ denotes equality up to exact cocycles and $H^\wedge_\cup$ is a homotopy between the cup and wedge products, see Appendix \ref{AppDiffCohom}. $\wedge_\Lambda$ and $\wedge_\mathbb{Z}$ denotes respectively the wedge products obtained from the pairing on $\Lambda$ and the standard unimodular pairing on $\mathbb{Z}$. We therefore recover up to an exact term the familiar Lagrangian $k C \wedge dC$. \eqref{EqLagWuCSTheory} generalizes it to topologically non-trivial fields, arbitrary abelian groups and half-integer levels. 

\paragraph{Action} The Lagrangian above has the puzzling feature that it is not gauge invariant modulo integers under large gauge transformations \cite{Monnier:2016jlo}, unless the lattice pairing is valued in $2\mathbb{Z}$, in which case we are dealing with ordinary abelian Chern-Simons theory. One cannot construct a gauge invariant action by simply integrating $L(\check{C})$ over the spacetime manifold $M$. The solution to this puzzle is that $L(\check{C})$ and $g_2 := g \mbox{ mod } 2$ define together a class $[L(\check{C}),g_2]$ in a certain generalized cohomology, named E-theory. This class is invariant under the gauge transformations of $\check{C}$. One can use the integration map in E-theory on this class to construct a gauge invariant action from the Lagrangian above \cite{Monnier:2016jlo}. The action therefore reads
\be
\label{EqDefActWuCSThFlat}
S_{\rm WCS}(M,\Lambda,\check{C}) = \frac{1}{2} \int_M^{\rm E} \left[ L(\Lambda,\check{C}), g_2 \right] \;,
\ee 
where we denoted the integration map in E-theory by $\int^{\rm E}$.

If the 7-dimensional $(2,0)$-manifold $M$ is bounded by an 8-dimensional $(2,0)$ manifold $W$, the action can be expressed as an ordinary integral of differential forms over $W$ \cite{Monnier:2016jlo}. As the inclusion of $M$ in $W$ is a morphism of $(2,0)$-manifolds, $\check{C}$ extends to $W$ as a differential cocycle $\check{C}_W$ shifted by the Wu class. We can extend as well the differential cocycle $\check{\nu}$ to a differential cocycle $\check{\nu}_W$ on $W$ whose characteristic $\nu_{\Lambda,W} := \nu_{\mathbb{Z},W} \otimes c$ lifts the Wu class, i.e. the periods of the $\mathbb{Z}$-valued cocycle $\nu_{\mathbb{Z},W}$ are even or odd depending on whether the periods of the Wu class of $W$ are 0 or 1. The field strength $\lambda_W$ of $\check{\nu}_W$ is then a differential form vanishing on $M$ lifting the Wu class.  Therefore 
\be
\label{EqDefC'}
\check{C}'_W := \check{C}_W - \frac{1}{2}\check{\nu}_W \;,
\ee
is an unshifted differential cocycle, with field strength $G'_W := G_W - \frac{1}{2}\lambda_W$. The action then reads
\be
\label{EqActWCSBndy}
S_{\rm WCS}(M,\Lambda,\check{C}) = \int_W  \left(\frac{1}{2} G_W \wedge G_W - \frac{1}{8} \lambda_W^2 \right) = \frac{1}{2} \int_W G'_W \wedge (G'_W + \lambda_W) \;.
\ee

\paragraph{Prequantum theory} Given the action, there is a standard way to construct from it an invertible field theory, the \emph{prequantum theory} associated to the action \cite{Freed:1991bn}. The partition function of the prequantum theory on a 7-dimensional $(2,0)$-manifold $M$ is simply the exponentiated action $\exp 2\pi i S_{\rm WCS}(M,\Lambda,\check{C})$. As is well-known in the case of ordinary Chern-Simons theory, on a manifold with boundary, the exponentiated action is not canonically a complex number: it is not gauge invariant, so its value as a complex number depends on a choice of gauge. However, it is possible to see it canonically as an element of a Hermitian line associated to the boundary (i.e. a 1-dimensional Hilbert space non-canonically isomorphic to $\mathbb{C}$). A choice of gauge determines an isomorphism with $\mathbb{C}$, thereby allowing to identify the exponentiated action with a complex number. This Hermitian line is the state space that the prequantum theory assigns to the 6-dimensional boundary. One can show that this data combines into a field theory functor in the sense of Atiyah-Segal, from the bordism category of 6-dimensional $(2,0)$-manifolds into the category of Hilbert spaces. We will write it
\be
\textcal{WCS}_{\rm P}[\Lambda, \check{C}] \;.
\ee
Despite the notation, the theory depends only on the differential cohomology class of the gauge field $\check{C}$. A detailed construction of $\textcal{WCS}_{\rm P}[\Lambda, \check{C}]$ can be found in Section 5 of \cite{Monnier:2016jlo}.

\subsection{Discrete gauging}

\label{SecDGWCS}

\paragraph{Symmetry} It was shown in \cite{Monnier:2016jlo} that there is an action of $H^3(M;\Lambda^\ast/\Lambda)$ on the group $\check{H}^4(M;\Lambda)$ of gauge equivalence classes of gauge fields on $M$. Up to possible anomalies, this action is a symmetry of the prequantum theory, and can therefore be gauged.

We refer the reader to Section 3 of \cite{Monnier:2016jlo} for a detailed description of this symmetry, but we can understand it as follows. There is a subgroup $\mathsf{C}$ of $H^3(M;\Lambda^\ast/\Lambda)$ consisting of classes $u$ which are reductions of classes in $H^3(M;\Lambda^\ast) \simeq H^3(M;\mathbb{Z}) \otimes \Lambda^\ast$. The action of this subgroup on the gauge field $\check{C}$ is to add $\Lambda^\ast$-valued holonomy (or "Wilson line") along a 3-dimensional cycle Poincaré dual to the class in $H^3(M;\mathbb{Z})$ determined by $u$. The subgroup of $\mathsf{C}$ coming from torsion classes in $H^3(M;\mathbb{Z})$ acts trivially. The quotient $\mathsf{K} = H^3(M;\Lambda^\ast/\Lambda)/\mathsf{C}$ is in bijection with a subgroup of $H_{\rm tors}^4(M; \Lambda)$ through the Bockstein map. The elements in $H^3(M;\Lambda^\ast/\Lambda)$ that project on non-trivial elements of $\mathsf{K}$ therefore add torsion fluxes in addition to holonomy. The fact that these operations are symmetries of the action essentially comes from the fact that the pairing between $\Lambda^\ast$ and $\Lambda$ is integer valued. Note also that if the lattice is unimodular, $\Lambda^\ast = \Lambda$ and the symmetry group above is trivial.

There is a convenient homomorphism from $H^3(M;\Lambda^\ast/\Lambda)$ into the differential cohomology group $\check{H}^4(M;\Lambda)$ that makes the action above obvious. Let $e$ be a cocycle representative of a class in $H^3(M;\Lambda^\ast/\Lambda)$. We can lift $e$ to a $\Lambda^\ast$-valued cochain $e_{\Lambda^\ast}$. Then 
\be
\label{EqDiffCocFromCoc}
\check{e} = (-de_{\Lambda^\ast}, e_{\Lambda^\ast}, 0)
\ee
is a $\Lambda$-valued differential cocycle defining a class in $\check{H}^4(M;\Lambda)$. The action of $H^3(M;\Lambda^\ast/\Lambda)$ on $\check{H}^4(M;\Lambda)$ is then just given by the addition of differential cohomology classes through this homomorphism.

\paragraph{Anomalies} The statements above should be qualified. Strictly speaking the action of $H^3(M;\Lambda^\ast/\Lambda)$ may change the sign of the exponentiated action. We can see this sign as an anomaly of the would-be symmetry. The action of $H^3(M;\Lambda^\ast/\Lambda)$ is a true symmetry, and not just a symmetry up to signs, only for certain choices of Wu structures (called \emph{admissible}), and for certain choices of torsion fluxes for $\check{C}$, as explained in Sections 4.5 and 4.6 of \cite{Monnier:2016jlo}. We will always assume that the Wu structure and fluxes have been chosen so that the action of $H^3(M;\Lambda^\ast/\Lambda)$ is a symmetry. 

\paragraph{The partition function of the gauged theory} The partition function of the gauged theory is up to a normalization factor the sum of the partition function of the prequantum theory, the exponentiated action, over the orbit of the action of $H^3(M;\Lambda^\ast/\Lambda)$. As the action itself is constant along the orbit, the only effect of the sum is to produce a prefactor. Combining it with the normalization factor, the partition function of the gauged theory reads:
\be
\label{EqPartFuncGaugWCS}
\textcal{WCS}_{\rm G}[\Lambda,\check{C}](M) = \mu_{M} \exp 2\pi i S_{\rm WCS}(M,\Lambda,\check{C}) \;,
\ee
where and
\be
\label{EqNormFactDGWCS}
\mu_{M} = \prod_{i = 0}^{3} |H^i(M;\Lambda^\ast/\Lambda)|^{(-1)^{3-i}} \;.
\ee 
See \cite{Monnier:2016jlo} for the case of a manifold with boundary.

\paragraph{The state space of the gauged theory} The construction of the state space of the gauged theory on a 6-manifold $N$ is quite subtle and is carried out in Sections 7 and 8 of \cite{Monnier:2016jlo}. It can be informally described as follows. 

There are Wilson operators associated to elements in $\mathsf{G} = H^3_{\rm free}(N;\mathbb{Z}) \otimes_\mathbb{Z} \Lambda^\ast/\Lambda$. They form a representation of the discrete Heisenberg group $\mathsf{H}$ associated to the skew symmetric non-degenerate $\mathbb{R}/\mathbb{Z}$-valued cup product pairing on $\mathsf{G}$. Let $V$ be the direct sum of the (one-dimensional) state spaces of the prequantum theory associated to the elements of the orbit of a gauge field $\check{C}$ on $N$ under the action of $H^3(N;\Lambda^\ast/\Lambda)$. Then $V$ decomposes into $|\mathsf{K}|$ copies of the regular representation of $\mathsf{H}$, where we recall that $\mathsf{K}$ is the image of $H^3(N;\Lambda^\ast/\Lambda)$ into $H^4_{\rm tors}(N;\Lambda)$ through the Bockstein homomorphism. 

The state space of the gauged theory is a certain quotient of the representation above, isomorphic to $|\mathsf{K}|$ copies of the irreducible representation of $\mathsf{H}$. Some extra data is actually required to identify the state space as a well-defined Hilbert space, due to the presence of Hamiltonian anomalies. The dimension of the Hilbert space is $|H^3(N;\Lambda^\ast/ \Lambda)|^{1/2}$.

We will describe the state space in a bit more detail in Section \ref{SecConfBlocks}.

\paragraph{Gluing conditions} The proof that the data above define a field theory functor, i.e. that it behaves consistently with the gluing of bordism, is far from straightforward and can be found in \cite{Monnier:2016jlo}.

\subsection{Wu Chern-Simons theories and lattice decompositions}

\label{AppDGWCSAndLatticeDec}

We now study the behavior of the field theories defined above under decompositions of the lattice $\Lambda$. These results did not appear in \cite{Monnier:2016jlo}.

\paragraph{Lattice decompositions} Suppose that we have a self-dual lattice $\Lambda$ into which we pick a sublattice of maximal dimension, which is itself decomposed into two orthogonal lattices $\Lambda_1$ and $\Lambda_2$. We want to understand how the gauged Wu Chern-Simons theories associated to the lattices $\Lambda$, $\Lambda_1$ and $\Lambda_2$ are related to each other. A typical example is the following. Let $\Lambda$ be the unit cubic lattice in three dimensions. Take $\Lambda_1$ to be the sublattice isometric to $\sqrt{3} \mathbb{Z}$ generated by $(1,1,1)$ and $\Lambda_2$ to be the $A_2$ sublattice given by the lattice elements in the plane orthogonal to $(1,1,1)$. $\Lambda_1 \oplus \Lambda_2$ is a sublattice of $\Lambda$ of maximal dimension and of index 3. More generally, we are interested in the case where $\Lambda$ is the $k$-dimensional unit cubic lattice, $\Lambda_1 = \sqrt{k} \mathbb{Z}$ and $\Lambda_2 = A_{k-1}$.

Let $V_i = \Lambda_i \otimes_\mathbb{Z} \mathbb{R}$. We can decompose any $a \in \Lambda$ as $a = a_1 + a_2$, $a_i \in V_i$. The integrality of the pairing on $\Lambda$ and the fact that $\Lambda_i \subset \Lambda$ imply that $a_i \in \Lambda_i^\ast$. Moreover, given $a_1 \in \Lambda_1^\ast$, the set of $a_2 \in \Lambda_2^\ast$ such that $a_1 + a_2 \in \Lambda$ forms a $\Lambda_2$-torsor. In fact, there is an isomorphism $\phi: \Lambda_1^\ast/\Lambda_1 \simeq \Lambda_2^\ast/\Lambda_2$ such that $a_1 + a_2 \in \Lambda$ if and only if the equivalence class of $a_2$ in $\Lambda_2^\ast/\Lambda_2$ is the image through $\phi$ of the equivalence class of $a_1$ in $\Lambda_1^\ast/\Lambda_1$.

The isomorphism $\phi$ induces an isomorphism between $H^3(M;\Lambda^\ast_1/\Lambda_1)$ and $H^3(M;\Lambda^\ast_2/\Lambda_2)$, which we write $\phi$ as well. Let us write $\mu_{M,\Lambda'}$, $\Lambda' = \Lambda, \Lambda_1, \Lambda_2$ for the normalization factor \eqref{EqNormFactDGWCS} of the discretely gauged theory. The isomorphism $\phi$ guarantees that $\mu_{M,\Lambda_1} = \mu_{M,\Lambda_2}$. As $\Lambda$ is self-dual, we obviously have $\mu_{M,\Lambda} = 1$.

\paragraph{Decomposition of the action} In the case of interest to us where $\Lambda$ is a unit square lattice, the element $c =(1,1,...,1)$ is characteristic: $(c,x) = (x,x)$ modulo 2 for all $x \in \Lambda$. It projects to characteristic elements of $\Lambda_1 = \sqrt{k} \mathbb{Z}$ and $\Lambda_2 = A_{k-1}$, namely $\sqrt{k}$ and $0$. (Recall that $A_{k-1}$ is an even lattice, so $0$ is a characteristic element.) We use $c$ to construct $\eta_\Lambda$ as in Section \ref{SecPreqWCS} and we have $\eta_{\sqrt{k}Z} = \eta_\Lambda$ and $\eta_{A_{k-1}} = 0$.

Let $\check{C}$ be a differential cocycle valued in $\sqrt{k}\mathbb{Z} \subset \Lambda$. As $\eta_{A_{k-1}} = 0$, we trivially have
\be
L(\Lambda,\check{C}) = L(\sqrt{k}\mathbb{Z},\check{C}) \;.
\ee
More generally, if $e$ is a cocycle representing a class in $H^3(M;\Lambda^\ast_1/\Lambda_1)$, the orthogonality of the lattices $\sqrt{k}\mathbb{Z}$ and $A_{k-1}$ imply
\be
L(\Lambda, \check{C} + \check{e} + \phi(\check{e})) = L(\sqrt{k}\mathbb{Z},\check{C} + \check{e}) + L(A_{k-1}, \phi(\check{e})) \;,
\ee
where $\check{e}$ is the differential cocycle constructed from $e$ as in \eqref{EqDiffCocFromCoc}. This equality holds at the level of the actions:
\be
\label{EqRelActLatDecomp}
S_{\rm WCS}(M,\Lambda, \check{C} + \check{e} + \phi(\check{e})) = S_{\rm WCS}(M,\sqrt{k}\mathbb{Z},\check{C} + \check{e}) + S_{\rm WCS}(M,A_{k-1},\phi(\check{e})) \;.
\ee

\paragraph{Relation between the prequantum theories} The relation \eqref{EqRelActLatDecomp} between the actions, valid as well on manifolds with boundary, implies immediately that the prequantum field theory functors are related by 
\be
\label{EqRelPreqFTFuncLatDec}
\textcal{WCS}_{\rm P}[\Lambda, \check{C}] \otimes \overline{\textcal{WCS}}_{\rm P}[\sqrt{k} \mathbb{Z}, \check{C}] = \textcal{WCS}_{\rm P}[A_{k-1}, 0] \;.
\ee
On a closed 7-dimensional $(2,0)$-manifold $M$, the tensor product sign in \eqref{EqRelPreqFTFuncLatDec} should be understood as multiplication and the bar as complex conjugation. \eqref{EqRelPreqFTFuncLatDec} is then a rephrasing of \eqref{EqRelActLatDecomp} for the exponentiated actions. On a closed 6-dimensional $(2,0)$-manifold $N$, $\otimes$ in \eqref{EqRelPreqFTFuncLatDec} denotes the tensor product and the bar is the complex conjugation of Hilbert spaces. \eqref{EqRelPreqFTFuncLatDec} holds as a result of the fact that the Hilbert space of the prequantum theory is constructed as the limit of a diagram of homomorphisms from $\mathbb{C}$ to $\mathbb{C}$ given by the exponentiated action. The equality of the actions implies directly a canonical isomorphism between the corresponding Hilbert spaces.

\paragraph{Relation between the gauged theories} The form of the partition function \eqref{EqPartFuncGaugWCS} shows that at the level of the gauged theories, we have
\be
\label{EqRelLatDecGaugedWCS}
\textcal{WCS}_{\rm G}[\Lambda,\check{C}](M) \otimes \overline{\textcal{WCS}_{\rm G}[\sqrt{k} \mathbb{Z},\check{C}](M)} = \textcal{WCS}_{\rm G}[A_{k-1},0](M) \;.
\ee
Indeed, as $\Lambda^\ast/\Lambda$ is the trivial group, the "gauged" Wu Chern-Simons theory associated to $\Lambda$ coincides with the prequantum theory. Moreover, the isomorphism $\phi$ ensures that the sums and normalization factors appearing in $\textcal{WCS}_{\rm G}[\sqrt{k} \mathbb{Z}, \check{C}](M)$ and $\textcal{WCS}_{\rm G}[A_{k-1}, 0](M)$ are the same. 

The state space of the gauged theory is constructed in two steps. First, a direct sum of the state space of the prequantum theory is taken over the orbit of the discrete gauge group. Then, a quotient is taken with respect the action of a certain groupoid, defined by the action of the prequantum theory on cylinders. The relation \eqref{EqRelPreqFTFuncLatDec} satisfied by the prequantum theory functor, applied throughout the construction above, implies that in dimension 6, we have as well:
\be
\label{EqRelLatDecGaugedWCSStSp}
\textcal{WCS}_{\rm G}[\Lambda,\check{C}](N) \otimes \overline{\textcal{WCS}_{\rm G}[\sqrt{k} \mathbb{Z},\check{C}](N)} = \textcal{WCS}_{\rm G}[A_{k-1},0](N) \;.
\ee

All in all we have the equality of the field theory functors
\be
\textcal{WCS}_{\rm G}[\Lambda, \check{C}] \otimes \overline{\textcal{WCS}_{\rm G}[\sqrt{k} \mathbb{Z}, \check{C}]} = \textcal{WCS}_{\rm G}[A_{k-1}, 0] \;.
\ee

\section{Anomaly field theory of a stack of M5-branes}

\label{SecAFTStack}

\subsection{M-theory backgrounds from (2,0)-structures}

\label{SecMThBack20Struct}

We explain here how the data of an $A_{k-1}$ $(2,0)$-structure on a $d$-dimensional manifold $U$ can be used to construct a "$(d+5)$-dimensional M-theory background". We will mostly be interested in $d = 6,7,8$, but the construction is independent of $d$. The case where the $(2,0)$ manifold is 7-dimensional and the M-theory background 12-dimensional is the one relevant to the computation of partition function anomalies, in keeping with the fact that partition function anomalies are described by geometric invariants of manifolds with one more dimension than the spacetime of the physical theory.

Consider the total space $E(\mathscr{N}_U)$ of the bundle $\mathscr{N}_U$. The connection and hermitian structure on $\mathscr{N}_U$, together with the Riemannian metric on $U$, yield a Riemannian metric on $E(\mathscr{N}_U)$. The spin structure on $TU \oplus \mathscr{N}_U$ yields a spin structure on $TE(\mathscr{N}_U)$. Pick $k$ sections of $\mathscr{N}_U$, to be seen as the worldvolumes of $k$ M5-branes. As a manifold, the M-theory background $Y$ is $E(\mathscr{N}_U)$ after the excision of the $k$ sections. They are excised because the M-theory C-field has a divergence on the worldvolumes. 

Let us now construct the C-field. Intuitively, the C-field is the one sourced by the $k$ M5-branes. The subtlety is that if $\mathscr{N}_U$ is non-trivial, there may not be a canonical C-field associated to the M5-brane configuration. In fact, we will only need the asymptotic value of the C-field on the complement of a tubular neighborhood of the stack, in the limit where the distances between the branes are scaled down to zero. Such a C-field can be constructed from the (2,0)-structure on $U$ as follows. Construct the differential cocycles $\check{b} := \frac{1}{2} \pi_\ast(\check{a} \cup \check{a})$ and $\check{A} := \check{C} - \check{b}$ on $U$. Note that $\check{A}$ is a differential cocycle shifted by $w_4(TU \oplus \mathscr{N}_U)$ \cite{Monnier:2014txa}. We require that on the boundary of a tubular neighborhood of radius $r$ enclosing all $k$ M5-brane, separated by a typical distance $r_{\rm stack}$, the C-field takes the form 
\be
\check{C}_{Y} = k\check{a} + \pi^\ast(\check{A}) + O(r_{\rm stack}/r) \;.
\ee
As $\check{a}$ is unshifted, $\check{C}_{Y}$ is a differential cocycle shifted by $w_4(TY)$, as required by membrane anomaly cancellation \cite{Witten:1996md}. We will not need to specify the C-field more explicitly in the following.

\subsection{Idea of the computation}

\label{SecDecompAn}

The partition function anomalies of a $d$-dimensional quantum field theory are described by a certain geometric invariant of $d+1$-dimensional manifolds, see for instance Section 2 of \cite{Monnier:2014txa}. This geometric invariant is supposed to be identified with the partition function of the anomaly field theory. The geometric invariant of 7-dimensional $(2,0)$-manifolds describing the anomaly of a stack of M5-branes can be computed through anomaly inflow, by evaluating the M-theory Chern-Simons term on a 4-sphere bundle $\tilde{U} \rightarrow U$ in $Y$ enclosing the stack of M5-branes \cite{Witten:1996hc, Harvey:1998bx, Monnier:2014txa}. Practically, we can use the fact that $\tilde{U}$ is the boundary of a 4-sphere bundle $\tilde{W}$ over the 8-manifold $W$ admitting $U$ as a boundary. This follows from the fact that the $(2,0)$-structure on $U$ extends to $W$, upon taking $\tilde{U}$ and $\tilde{W}$ to be the unit sphere bundles in $\mathscr{N}_U$ and $\mathscr{N}_W$, respectively. The anomaly is then given by the integral over $\tilde{W}$ of the 12-dimensional characteristic form ${\rm CS}_{12}$ associated to the M-theory Chern-Simons term:
\be
{\rm CS}_{12}(\tilde{W},\check{C}_{\tilde{W},k}) = 2\pi i \int_{\tilde{W}} \left(\frac{1}{6} G \wedge G \wedge G - G \wedge I_8 \right) \;,
\ee
where $G$ is the field strength of the C-field $\check{C}_{\tilde{W},k} := k \check{a}_{\tilde{W}} + \pi^\ast(\check{A}_W)$ on $\tilde{W}$, in the limit $r_{\rm stack}/r \rightarrow 0$. The index density $I_8$ is defined in terms of the Pontryagin forms of $T\tilde{W}$ by
\be
\label{EqDefI8}
I_8 = \frac{1}{48} \left(p_2(T\tilde{W}) - \left(\frac{p_1(T\tilde{W})}{2}\right)^2\right) \;.
\ee

Note that we do allow for configurations in which the M5-branes intersect. This is crucial to ensure that every 7-dimensional $(2,0)$-manifold subject to \eqref{EqConst20Mfld} is a boundary. The fact that the M5-branes intersect is irrelevant for the computation of the anomaly, as the latter is computed on a tubular neighborhood of the stack. Note however that strictly speaking, global anomaly cancellation in M-theory backgrounds containing M5-branes was checked only for non-intersecting M5-branes \cite{Monnierb}. We are assuming here that M-theory is anomaly free in the presence of configurations of intersecting M5-branes.

\subsection{Anomaly of the stack}

The integral of ${\rm CS}_{12}$ over the fibers of $\tilde{W}$ has been performed in \cite{Monnier:2014txa}, see also \cite{Harvey:1998bx}. We obtain
\be
\label{EqResIntMthCSTerm}
\frac{1}{2\pi i} \ln {\rm An}_{{\rm Stack},k}(U) = \int_{\tilde{W}} {\rm CS}_{12}(\tilde{W}, \check{C}_{\tilde{W},k}) = \int_W \left( kJ_8 - \frac{k^3-k}{24} p_2(\mathscr{N}_W) - \frac{k}{2}G_{W,k}^2 \right) \;,
\ee
where $G_{W,k}$ is the curvature of $\check{C}_{W,k} := k \check{b} + \check{A}$. $J_8$ is defined by 
\be
J_8 := I_8 - \frac{1}{24} p_2(\mathscr{N}_W)\;, 
\ee
where $I_8$ has is the same expression as in \eqref{EqDefI8}, but involves now the Pontryagin classes of $TW$. \eqref{EqResIntMthCSTerm} depends only on the $(2,0)$-structure of $U$ and has now to be interpreted as the partition function of the anomaly field theory of the stack. To this end, it is useful to reformulate it a bit. Let us write $\check{C}_W = \check{C}_{W,1}$ and $G_W$ for its curvature. We add and subtract $\frac{k}{2} G_W^2$ in the expression above and rearrange the terms as follows
\be
\label{EqAnomStackM5}
\begin{aligned}
\frac{1}{2\pi i} \ln {\rm An}_{{\rm Stack},k}(U) = \: & k \int_W \left( J_8 - \frac{1}{2} G_W^2 \right) \\
& + \int_W \left( - \frac{k^3-k}{24} p_2(\mathscr{N}_W) - \frac{k}{2}G_{W,k}^2 + \frac{k}{2}G_{W}^2 \right) \;.
\end{aligned}
\ee
The first term can be naturally interpreted as $k$ times the anomaly of a single $M5$-brane, while the second term is the anomaly due to the "Hopf-Wess-Zumino terms" of \cite{Intriligator:2000eq}. (See also Section 4.7 of \cite{Monnier:2014txa} for a geometrical interpretation of the Hopf-Wess-Zumino terms, in the case where the M5-branes do not intersect.) We will consider these terms separately.

\subsection{Anomaly of the M5-brane worldvolumes}

\label{SecAnM5WV}

Recall the first term of \eqref{EqAnomStackM5}:
\be
\label{EqAnnM53}
\frac{1}{2\pi i} \ln {\rm An}_{kM5}(U) = k \int_W \left( J_8 - \frac{1}{2}G_{W}^2 \right) \;,
\ee
which we now rewrite in a purely 7-dimensional form. We have \cite{Witten:1996hc}
\be
\label{ExIndDensFerm}
I_f = - 2J_8 + \frac{1}{4}L(TW) \;,
\ee
where $L(TW)$ is the degree 8 form component of the Hirzebruch genus of $TW$ and $I_f$ is the degree 8 index density of the Dirac operator $D_{f,W}$ on $W$ associated to the spinor bundle of $TW \oplus \mathscr{N}_W$. $I_f$ is twice the index density computing the local anomaly of the chiral fermions on a single M5-brane, the factor two being due to the symplectic Majorana condition satisfied by the latter. We write 
\be
\label{EqDecompGWLambWGpW}
G_W = \frac{1}{2} \lambda_W + G'_W
\ee
where $\lambda_W$ is a differential form representative of the cocycle determining the shift of $\check{C}_W$, as in \eqref{EqActWCSBndy}. $G'_W$ is a differential form with integral periods. We have
\begin{align}
\label{EqAnnM54}
\frac{1}{2\pi i} \ln {\rm An}_{M5}(U) = & -\frac{k}{2} \int_W I_f - \frac{k}{8} \int_W (\lambda_W^2 - L(TW)) - \frac{k}{2} \int_W G'_W(G'_W + \lambda_W) \;.
\end{align}

Let us analyse the terms of \eqref{EqAnnM54} one by one. The Atiyah-Patodi-Singer theorem \cite{Atiyah1973} expresses the integral of $I_f$ over $W$ in terms of the modified eta invariant \cite{MR861886, Dai:1994kq} $\xi_f(U)$ on $U$:
\be
\label{EqXifFromW}
\xi_f(U) = \int_W I_f - {\rm index}(D_{f,W})
\ee
$\xi_f(U)$ is given in terms of the ordinary eta invariant $\eta_f(U)$ of $D_{f,U}$ as
\be
\xi_f(U) = \frac{\eta_f(U) + h_U}{2}
\ee 
where $h_U$ is the dimension of the space of zero modes of $D_{f,U}$. As $D_{f,W}$ is quaternionic, ${\rm index}(D_{f,W})$ is even, so the first term of \eqref{EqAnnM54} can be rewritten as $-\frac{k}{2} \xi_f(U)$ mod 1. 

The second term is $k$ times the geometric invariant ${\rm hs}(U)$ described by Hopkins and Singer in \cite{hopkins-2005-70}. Note that it has an implicit dependence on the choice of Wu structure on $U$, as the latter determines the periods of $\lambda_W$.

The third term is the action of a Wu Chern-Simons theory on $U$
\be
k \times \frac{1}{2} \int_U^{\rm E} \check{C}'_U \cup (\check{C}'_U + \check{\nu}_U) \;,
\ee
where we used the same notation as in \eqref{EqDefC'}:
\be
\check{C}'_U = \check{C}_U - \frac{1}{2} \check{\nu}_U \;.
\ee
Let $\Lambda$ be the $k$-dimensional unit cubic lattice, with canonical basis $\{e_i\}$. We can rewrite the action above as the action of a Wu Chern-Simons theory on $U$ whose field $\check{C}^\Lambda_U = \sum_i e_i \check{C}'_U$ takes value in $\Lambda$ and is diagonal:
\be
\label{EqActWCSLatLambda}
S_{\rm WCS}(U, \Lambda, \check{C}^\Lambda_U) = \frac{1}{2} \int_U^{\rm E} \check{C}^\Lambda_U \cup (\check{C}^\Lambda_U + \check{\nu}^\Lambda_U) \;,
\ee
where $\check{\nu}^\Lambda_U = \sum_i e_i \check{\nu}_U$. The $i$th component of $\check{C}^\Lambda_U$ is interpreted as the effective M-theory C-field on the worldvolume of the $i$th M5-brane. $\check{C}^\Lambda_U$ is diagonal because in the limit $r_{\rm stack} \rightarrow 0$, the effective C-field is the same on each M5-brane. Like the Hopkins-Singer term, \eqref{EqActWCSLatLambda} has a dependence on the Wu structure. As \eqref{EqAnnM53} had no dependence on the Wu structure, it exactly cancels against the Wu structure dependence of the Hopkins-Singer term.

We conclude that the anomaly of the worldvolume theory of $k$ M5-branes can be rewritten as 
\begin{align}
\label{EqAnnM55}
\frac{1}{2\pi i} \ln {\rm An}_{kM5}(U) = & -\frac{k}{2} \xi_f(U) - k{\rm hs}(U) - S_{\rm WCS}(U, \Lambda, \check{C}^{\Lambda}_U) \;.
\end{align}

\subsection{Anomaly field theory of the M5-brane worldvolumes}

\label{SecAFTM5WV}

The form of the anomaly \eqref{EqAnnM55} suggests that the anomaly field theory of the stack of M5-branes is given by
\be
\label{EqAFTTrivStackM5}
\textcal{An}_{kM5} = \left(\textcal{DF}^{\frac{1}{2}}_{f} \right)^{\otimes (-k)} \otimes \left( \textcal{HS} \right)^{\otimes (-k)} \otimes \overline{\textcal{WCS}}_{\rm G}[\bar{\Lambda}, \check{C}^{\bar{\Lambda}}] \;,
\ee
where each factor represents a quantum field theory, as defined below. The tensor product denotes the tensor product of the associated functors, corresponding physically to taking non-interacting copies of the quantum field theories to live on the same spacetime. $(\bullet)^{\otimes k}$ means taking the $k$th tensor product of the theory within the brackets. Negative exponents make sense for invertible field theories. The bar over the symbol of a quantum field theory denotes complex conjugation. For unitary invertible theories, complex conjugation is equivalent to $(\bullet)^{\otimes -1}$.

\paragraph{Half Dai-Freed theory} Given a Dirac operator on $d$-dimensional manifolds, for $d$ odd, Dai and Freed constructed in \cite{Dai:1994kq} a $d$-dimensional field theory functor $\textcal{DF}$. The partition function of $\textcal{DF}$ on a $d$-dimensional manifold $M$ is the exponential of the modified eta invariant $\xi$:
\be
\textcal{DF}(M) = \exp 2\pi i \xi \;.
\ee
The state space assigned by $\textcal{DF}$ to a $d-1$-dimensional manifold is the determinant line of the chiral Dirac operator obtained from the restriction of the $d$-dimensional Dirac operator. It was shown in \cite{Dai:1994kq} that the exponentiated modified eta invariant on a manifold with boundary takes value in the determinant line of the boundary, and that this data glue consistently. The extension of this field theory as an extended field theory down to codimension 2 was sketched in \cite{Monnierd}.

The field theory $\textcal{DF}^{\frac{1}{2}}_{f}$ appearing as the first factor of \eqref{EqAFTTrivStackM5} is not quite a conventional Dai-Freed theory. It is a "half" or a "square root" of the Dai-Freed theory associated to the Dirac operator $D_f$, due to the fact that it acts on spinors obeying a symplectic Majorana condition, which effectively divides by two the number of degrees of freedom. It can be described as follows.

On a 7-dimensional $(2,0)$-manifold $U$, the partition function is defined as 
\be
\label{EqSqrtModEtaInv}
\textcal{DF}^{\frac{1}{2}}_{f}(U) := \exp \pi i \xi_f \;,
\ee
where $\xi_f$ is as above the modified eta invariant associated to the Dirac operator $D_{f,U}$. There is a potential ambiguity in the expression \eqref{EqSqrtModEtaInv}, because as the Dirac operator may experience spectral flow as the background data is changed, $\xi_f$ is a priori defined only modulo 1. However, the fact that $D_{f,U}$ is quaternionic shows that $\xi_f$ may only jump by even integers when the background data is varied, thereby ensuring that \eqref{EqSqrtModEtaInv} is a well-defined function over the space of background data. 

On a 6-dimensional $(2,0)$-manifold $M$, the fact that $D_{f,M}$ is quaternionic ensures that $D_{f,M}$ admits a Pfaffian line ${\rm Pfaff}(D_{f,M})$ whose square is the determinant line of $D_{f_M}$. The state space that $\textcal{DF}^{\frac{1}{2}}_{f}$ assigns to $M$ is the Hermitian line ${\rm Pfaff}(D_{f,M})$. More details about the construction of the Pfaffian line can for instance be found in Section 1.3 of \cite{Freed:2004yc}.

The proof of the gluing relations showing that $\textcal{DF}^{\frac{1}{2}}_{f}$ is a field theory functor should follow the same lines as the original analysis of Dai and Freed \cite{Dai:1994kq}, although to our knowledge this has not been worked out in detail.

\paragraph{Hopkins-Singer theory} The second factor $\textcal{HS}$ in \eqref{EqAFTTrivStackM5} denotes an invertible quantum field theory that can be straightforwardly constructed from the results of Hopkins and Singer in \cite{hopkins-2005-70}. In fact, we will see that this factor cancels against an identical factor coming from the center of mass of the stack of M5-branes, and hence will not appear in the anomaly field theory of the $(2,0)$ SCFTs. We will therefore not describe $\textcal{HS}$ in detail.

In short, the proof of the main theorem of \cite{hopkins-2005-70} involves the construction of a map from a Thom spectrum into a spectrum known as the Anderson dual of the sphere. But one can associate to such a map an invertible field theory functor (see for instance \cite{2014arXiv1406.7278F}). Maybe more concretely, one can specialize the main theorem of \cite{hopkins-2005-70} to the case where $S$ is a point and $k = 2$ in their notation. It then states that to each $8-p$-dimensional manifold endowed with an integral lift of the Wu class and a differential cocycle, one can associate a differential cocycle of degree $p$ over a point. This assignment is compatible with the gauge transformations of differential cocycle and with the gluing of manifolds. But a differential cocycle of degree 1 over a point is just an element of $\mathbb{R}/\mathbb{Z}$, so their theorem assign and element of $\mathbb{R}/\mathbb{Z}$ to a 7-manifold, which is to be interpreted as the log of the partition function of the Hopkins-Singer field theory. Similarly, a differential cocycle of degree 2 over a point is a Hermitian line, to be interpreted as the state space that the Hopkins-Singer field theory assigns to a 6-manifold. Their construction implicitly specifies a fully extended field theory, as $p$ can be as large as $8$.

\paragraph{Wu Chern-Simons theory} The last term in \eqref{EqAFTTrivStackM5} is the Wu Chern-Simons theory associated to the $k$-dimensional cubic lattice $\Lambda$, described in Section \ref{SecPreqWCS} and constructed in detail in Section 5 of \cite{Monnier:2016jlo}. Note that as $\Lambda$ is self-dual, there is no difference between the prequantum Wu Chern-Simons theory and the discretely gauged one (the discrete gauge group is trivial). We choose here to see it as a gauged theory.

\subsection{Hopf-Wess-Zumino anomaly}

\label{SecHWZAnomaly}

We now turn to the second term in \eqref{EqAnomStackM5}:
\be
\frac{1}{2\pi i} {\rm An}_{\rm HWZ}(U) = \int_W \left( - \frac{k^3-k}{24} p_2(\mathscr{N}_W) - \frac{k}{2}G_{W,k}^2 + \frac{k}{2}G_{W}^2 \right) \;.
\ee
Writing $h_W$ for the field strength of $\check{b}_W$ and using $G_{W,k} = G_W + (k-1)h_W$, this can be rewritten as the sum of two terms \cite{Monnier:2014txa}
\be
\frac{1}{2\pi i} {\rm An}_{\rm HWZ}(U) = S_{\rm HWZ}^{(3)} + S_{\rm HWZ}^{(2)} \;,
\ee
with
\be
\label{EqCubHWZTerm}
S_{\rm HWZ}^{(3)}(U) = -(k^3-k) \int_W \left( \frac{1}{24} p_2(\mathscr{N}_W) + \frac{1}{2} h_W^2 \right) \;,
\ee
\be
\label{EqQuadHWZTerm}
S_{\rm HWZ}^{(2)}(U) =  - k(k-1) \int_W h_W \left( G_W - h_W \right) \;.
\ee
It was explained in \cite{Monnier:2014txa} why these two terms are integers for closed $W$, and therefore well-defined geometric invariants modulo 1 associated to $U$. We now reexpress \eqref{EqCubHWZTerm} and \eqref{EqQuadHWZTerm} in terms of quantities defined on $U$.

\paragraph{The term \eqref{EqCubHWZTerm}} It is easy to check that $k^3-k$ is always a multiple of 6, so we can rewrite
\be
S_{\rm HWZ}^{(3)}(U) = -\frac{k^3-k}{6} \int_W \left( \frac{1}{4} p_2(\mathscr{N}_W) + 3 h_W^2 \right) \;,
\ee
where the prefactor is an integer. Recall that $h_W$ is the curvature of $\check{b}_W$, which is a differential cocycle shifted by $w_4(\mathscr{N}_W)$. Writing $b_W$ for the characteristic of $\check{b}_W$, this means that $2b_W$ is an integer-valued cocycle lift of $w_4(\mathscr{N}_W)$, or equivalently that $2b_W$ modulo 2 is a cocycle representative of $w_4(\mathscr{N}_W)$. We showed in \cite{Monnier:2014txa} that any such cocycle lift satisfies 
\be
\label{EqRelbp2Cohom}
[(2b_W)^2] = [p_{2,W}] \quad \mbox{ mod } 4\;,
\ee
where $p_{2,W}$ is any integral cocycle representative of the second Pontryagin class, and the bracket denote the ($\mathbb{Z}$-valued) cohomology class.

\eqref{EqRelbp2Cohom} can be promoted to an equality of cocycles as follows. Recall from Appendix \ref{AppEulerStruct} that $BSO[e](5)$ is the homotopy fiber of the classifying map of the Euler class of the universal bundle over $BSO(5)$. $BSO[e](5)$ carries a rank 5 bundle $\mathscr{N}$ given by the pullback of the universal bundle over $BSO(5)$, and by definition, the Euler class of $\mathscr{N}$ vanishes. As explained in Appendix \ref{AppEulerStruct}, we choose a degree 4 cocycle $a$ on the unit sphere bundle $\widetilde{BSO}[e](5)$ of $\mathscr{N}$ that represents the top cohomology class on each fiber. The Euler structure in the $(2,0)$-structure of $W$ provides a lift to $BSO[e](5)$ of the classifying map of $\mathscr{N}_W$ into $BSO(5)$, and $\mathscr{N}_W$ coincides with the pullback of $\mathscr{N}$ through this lift. $b_W$ is the pull-back of the $\frac{1}{2} \mathbb{Z}$-valued cocycle $b := \frac{1}{2}\pi_\ast(a \cup a)$. The discussion of \cite{Monnier:2014txa} mentioned the previous paragraph applies to $\mathscr{N}$, showing that $(2b)^2$ represents the second Pontryagin class of $\mathscr{N}$ modulo 4. We can therefore choose a cocycle $p_{2}$ on $BSO[e](5)$ representing the second Pontryagin class of $\mathscr{N}$ such that 
\be
\label{EqRelp2bUniv}
p_{2} = (2b)^2 \quad \mbox{mod } 4 \;.
\ee
The choice of universal cocycles $a$, $b$ and $p_{2}$ on $BSO[e](5)$ induces corresponding cocycles satisfying \eqref{EqRelp2bUniv} on every $(2,0)$-manifold, in a way compatible with the morphisms of $(2,0)$-manifolds. In particular, $p_{2}$ induces via pull-back an integral cocycle $p_{2,W}$ on $W$ representing the second Pontryagin class of $\mathscr{N}_W$ and satisfying $p_{2,W} = 2b_W$ mod 4. Then the combination 
\be
\frac{1}{4} p_{2,W} + 3 b_W^2
\ee
is an integer-valued cocycle.

We can use $p_{2,W}$ and the Riemannian metric on $W$ to construct a canonical differential cocycle representative $\check{p}_2(\mathscr{N}_W)$, whose characteristic is $p_{2,W}$ and whose curvature is the second Pontryagin form $p_2(\mathscr{N}_W)$ (see Theorem 2.2 of \cite{springerlink:10.1007/BFb0075216}). We can rewrite the integrand of \eqref{EqCubHWZTerm} as the curvature of a differential cocycle:
\be
S^{(3)}_{\rm HWZ} = -\frac{k^3-k}{6} \int_W \left[ \frac{1}{4}\check{p}_2(\mathscr{N}_W) + 3 \check{b}^2_W \right]_{\rm cu} \;.
\ee 
(Note that we have defined the cup product in Appendix \ref{AppDiffCohom} only on unshifted differential cocycles. The integrand above should be read $\frac{1}{4}\left(\check{p}_2(\mathscr{N}_W) + 3 (2\check{b}_W)^2 \right)$, which indeed involves only unshifted differential cocycles.) Using the closure relation $[\check{c}]_\omega = [\check{c}]_{\rm ch} + d[\check{c}]_{\rm co}$ valid for any differential cocycle $\check{c}$, we obtain
\be
S^{(3)}_{\rm HWZ}(U) = -\frac{k^3-k}{6} \int_W \left( \left[ \frac{1}{4}\check{p}_2(\mathscr{N}_W) + 3 \check{b}^2_W \right]_{\rm ch} + d\left[ \frac{1}{4}\check{p}_2(\mathscr{N}_W) + 3 \check{b}^2_W \right]_{\rm co} \right)\;.
\ee 
But as noted above, the first term is an integer, and the second one can be rewritten as an integral over $U$:
\be
S^{(3)}_{\rm HWZ}(U) = -\frac{k^3-k}{6} \int_U \left[ \frac{1}{4}\check{p}_2(\mathscr{N}_U) + 3 \check{b}^2_U \right]_{\rm co} \mbox{ mod } 1\;.
\ee 
$\check{p}_2(\mathscr{N}_U)$ is the differential cocycle representative of the second Pontryagin class of $\mathscr{N}_U$ constructed from the Riemannian metric on $U$ and from the integral cocycle $p_{2,U}$ pulled back from $F$. It coincides with the restriction of $\check{p}_2(\mathscr{N}_W)$ to $U$ by the discussion above. This provides an expression for $S^{(3)}_{\rm HWZ}(U)$ in terms of quantities defined on $U$ only. We will write 
\be
\label{EqActCSp2Th}
S_{{\rm CS}p_2} := \int_U \left[ \frac{1}{4}\check{p}_2(\mathscr{N}_U) + 3 \check{b}^2_U \right]_{\rm co} \;.
\ee

\paragraph{The term \eqref{EqQuadHWZTerm}} Using \eqref{EqDecompGWLambWGpW} and the fact that $k(k-1)$ is even, we can rewrite
\be
\label{EqQuadHWZTerm2}
S_{\rm HWZ}^{(2)}(U) =  \frac{k(k-1)}{2} \left( \frac{1}{2} \int_W (-2h_W) \left( \lambda_W + (-2h_W) \right) + \int_W (-2h_W) G'_W \right) \;,
\ee 
where the prefactor is an integer, and $-2h_W$ and $G'_W$ are differential forms with integer periods. Comparing with \eqref{EqActWCSBndy}, we see that \eqref{EqQuadHWZTerm2} is the sum of the action of a Wu Chern-Simons theory and of a BF theory. The BF theory action can easily be expressed on $U$ using the cup product of differential cocycles, so we obtain
\be
\label{EqQuadHWZTerm3}
S_{\rm HWZ}^{(2)}(U) =  \frac{k(k-1)}{2} \left( S_{WCS}(U, \mathbb{Z}, -2\check{b}_U) + S_{\rm BF}(U, -2\check{b}_U, \check{C}_U' ) \right) \;,
\ee 
\be
\label{EqActBFTh}
S_{\rm BF}(U, -2\check{b}_U, \check{C}_U') := \int_U \left[ (-2\check{b}_U) \cup \check{C}'_U \right]_{\rm co}
\ee

\subsection{Hopf-Wess-Zumino anomaly field theory}

\label{SecAFTHWZ}

The form of the anomaly above suggests the following anomaly field theory. 
\be
\label{EqAFTHWZ}
\textcal{An}_{\rm HWZ} =  \left( \textcal{WCS}_{\rm P}[\mathbb{Z}, -2\check{b}]\right)^{\otimes \frac{k(k-1)}{2}} \otimes
\left(\textcal{BF}[-2\check{b}, \check{C}']\right)^{\otimes \frac{k(k-1)}{2}} \otimes 
\left( \textcal{CSp}_2[\check{b}] \right)^{\otimes \frac{k - k^3}{6}} \;,
\ee
where the factors are detailed below.

\paragraph{Prequantum Wu Chern-Simons theory} The first factor $\textcal{WCS}_{\rm P}[\mathbb{Z}, -2\check{b}]$ is a prequantum Wu Chern-Simons theory associated to the lattice $\mathbb{Z}$, with background gauge field given by $-2\check{b}$, see Section \ref{SecPreqWCS}.

\paragraph{Abelian prequantum BF theory} The second factor $\textcal{BF}[-2\check{b}, \check{C}']$ is the prequantum abelian BF theory with action \eqref{EqActBFTh}. This is an invertible quantum field theory that can be constructed along the lines of Section \ref{SecPreqWCS} (see also \cite{Freed:1991bn} or Section 4 of \cite{Monnierd}). 

In short, the partition function of the prequantum abelian BF theory on a 7-manifold is simply the exponentiated action. On a 7-manifold with boundary, because of the failure of gauge invariance, the exponentiated action is not quite a complex number, but rather an element of a Hermitian line associated to the 6-dimensional boundary. This hermitian line is the state space of the prequantum abelian BF theory on the 6-dimensional boundary. More abstractly, the state space can be constructed by "integrating the degree 7 cocycle over the 6-dimensional boundary", a procedure introduced in \cite{Freed:1991bn} in the context of Dijkgraaf-Witten theory.

\paragraph{Abelian prequantum Chern-Simons-$p_2$ theory} For lack of a better name, we call the third factor $\textcal{CSp}_2[\check{b}]$ a Chern-Simons-$p_2$ theory. In the action \eqref{EqActCSp2Th}, the second term looks like a usual quadratic Chern-Simons term for the $U(1)$ 3-form gauge field associated to the differential cocycle $\check{b}_U$. The subtlety is that this gauge field is shifted by $w_4(U)$: its fluxes on four-cycles are integral or half-integral depending on whether the periods of $w_4(U)$ are zero or one modulo 2. Such a quadratic term would not be well-defined on its own, but the presence of the first term proportional to the second Pontryagin class makes the whole action well-defined, as explained in Section \ref{SecHWZAnomaly}. 

As the Lagrangian $\frac{1}{24}\check{p}_2(\mathscr{N}_U) + \frac{1}{2} \check{b}^2_U$ is an unshifted differential cocycle of degree 8, the procedure sketched above in the case of the BF action yields an invertible quantum field theory, the prequantum Chern-Simons-$p_2$ theory $\textcal{CSp}_2[\check{b}]$.

\paragraph{Dependence on the Wu structure} $\textcal{WCS}_{\rm P}[\mathbb{Z}, -2\check{b}]$ depends on the Wu structure of the $(2,0)$ manifold, and so does $\textcal{BF}[-2\check{b}, \check{C}']$, through its argument $\check{C}'$. However, \eqref{EqQuadHWZTerm} makes it clear that the product theory depends only on $\check{b}$ and $\check{C}$, and is therefore independent of the Wu structure.

\section{Anomaly field theory of the center of mass}

\label{SecAFTCM}

The $(2,0)$ $A_{k-1}$ theory is obtained from the worldwolume theory of a stack of $k$ M5-branes by removing the center of mass degrees of freedom, associated to the collective excitations of the M5-branes. The center of mass degrees of freedom form a free tensor multiplet carrying a charge $k$ with respect to the background C-field. The subtle part of the anomaly of the tensor multiplet comes from the anomaly of the charge $k$ self-dual field it contains. In this section, we first analyse the anomalies and anomaly field theory of the charge $k$ self-dual field and then those of the charge $k$ tensor multiplet.

\subsection{Anomaly of a charge $k$ self-dual field}

\label{SecAnChgkSDF}

Consider a degree 2 self-dual gauge field in six dimension. This theory can be naturally coupled to a degree 3 $U(1)$ gauge field \cite{Witten:1996hc}. The self-dual field can have arbitrary integer charge $k$ with respect to this $U(1)$ gauge field \cite{Belov:2006jd}. The anomalies of the charge $k$ self-dual field are encoded in a geometric invariant ${\rm An}_{SD(k)}$ of 7-dimensional smooth oriented Riemannian manifolds endowed with a degree 3 $U(1)$ gauge field. For our purpose, it will be sufficient to restrict ourselves to the case where the 7-dimensional manifold $U$ and the gauge field $\check{C}_U$ are obtained from a $(2,0)$-manifold by forgetting the bundle $\mathscr{N}$. The results of Appendix \ref{SecCobGroup20Struct} show that there is always an 8-dimensional $(2,0)$-manifold $W$ bounded by $U$ endowed with a $U(1)$ gauge field $\check{C}_W$ shifted by the Wu class and satisfying $\check{C}_W|_U = \check{C}_U$. As explained in Section 4.3 of \cite{Monnier:2014txa}, we have
\be
\label{EqAnSDk1}
\frac{1}{2\pi i} \ln {\rm An}_{{\rm SD}(k)}(U,\check{C}_U) = \frac{1}{8} \eta_\sigma(U) -  k \int_W \left(\frac{1}{2}G^2_W - \frac{1}{8}\sigma_W \right) \;,
\ee
where $\eta_\sigma(U)$ is the eta invariant of the signature Dirac operator on $U$, $\sigma_W$ is the signature of the wedge product pairing on the cohomology of $W$ relative to $U$ and $G_W$ is the field strength of $\check{C}_W$.

We now give a purely 7-dimensional formula for the anomaly above. The Atiyah-Patori-Singer theorem \cite{Atiyah1973} allows us to write the signature as
\be
\sigma_W = \int_W L(TW) - \eta_\sigma(U) \;.
\ee
Using \eqref{EqDecompGWLambWGpW}, we have 
\be
\label{EqAnSDk}
\frac{k}{2} \int_W G^2_W - \frac{k}{8} \sigma_W(U) = \frac{k}{8} \eta_\sigma + \frac{k}{8} \int_W \left(  \lambda_W^2 - L(TW) \right) + \frac{k}{2} \int_W G'_W \wedge (G'_W + \lambda_W)
\ee
Comparing with \eqref{EqActWCSBndy}, we see that the third term coincides with minus the action of the Wu Chern-Simons theory on $U$ associated to the lattice $\sqrt{k}\mathbb{Z}$, which is the charge lattice of the self-dual field of charge $k$. The second term is $k$ times the geometric invariant described by Hopkins and Singer in \cite{hopkins-2005-70}, which already appeared in \eqref{EqAnnM55}. We therefore have
\be
\label{EqAnSDk2}
\frac{1}{2\pi i} \ln {\rm An}_{{\rm SD}(k)}(U,\check{C}_U) = -\frac{k-1}{8} \mathrm{\eta}_\sigma(U) -  k {\rm hs}(U)  - S_{WCS}(U,\sqrt{k}\mathbb{Z},\check{C}_U) \;.
\ee
\eqref{EqAnSDk2} provides a manifestly 7-dimensional expression for the anomaly of a charge $k$ self-dual field.

\subsection{Anomaly field theory of the charge $k$ self-dual field}

\label{SecAFTSDK}

Our task is now to interpret \eqref{EqAnSDk2} as the partition function of a quantum field theory, which we would identify with the anomaly field theory of a self-dual field of charge $k$. \eqref{EqAnSDk2} suggests that the relevant field theory is
\be
\label{EqAFTSDk}
\textcal{An}_{{\rm SD}(k)} = \left( \textcal{DF}^{\frac{1}{4}}_\sigma \right)^{\otimes (-k+1)} \otimes \textcal{HS}^{\otimes (-k)} \otimes  \overline{\textcal{WCS}}_{\rm G}\left[\sqrt{k}\mathbb{Z}\right]
\ee
We now describe the quantum field theories appearing in \eqref{EqAFTSDk} and comment on their relation to \eqref{EqAnSDk2}. 

\paragraph{Quarter Dai-Freed theory} $\textcal{DF}^{\frac{1}{4}}_\sigma$ is a "fourth root" of the Dai-Freed theory (see Section \ref{SecAFTM5WV}) associated to the signature Dirac operator. Recall that the signature Dirac operator \cite{Atiyah1973} on $U$ is constructed from the differential and the Hodge star operator. Its kernel can be expressed in terms of the cohomology of $U$, and has therefore constant rank over the connected components of the moduli space of 7-dimensional $(2,0)$-manifolds. This means that unlike the eta invariants associated to other Dirac operators, $\eta_\sigma(U)$ is well-defined as a real number, not only as a real number modulo 2. Therefore $\frac{1}{8} \eta_\sigma(U)$ is well-defined. The partition function of $\textcal{DF}^{\frac{1}{4}}_\sigma$ on $U$ is
\be
\label{EqPartFuncDF14}
\textcal{DF}^{\frac{1}{4}}_\sigma(U) = \exp \left(2 \pi i \frac{1}{8} \eta_\sigma(U) \right) \;.
\ee

The standard construction of a prequantum field theory, already sketched in Sections \ref{SecPreqWCS} and \ref{SecAFTHWZ}, combined with the definition \cite{Dai:1994kq} of the eta invariant on manifolds with boundary, can be applied to the exponentiated action \eqref{EqPartFuncDF14} to yield an invertible field theory $\textcal{DF}^{\frac{1}{4}}_\sigma$. The proof of the gluing axioms should follow from the corresponding proof in \cite{Dai:1994kq}. 

\paragraph{Hopkins-Singer theory} The Hopkins-Singer theory $\textcal{HS}$ has already been discussed in Section \ref{SecAFTM5WV}.

\paragraph{Discretely gauged Wu Chern-Simons theory} $\overline{\textcal{WCS}}_{\rm G}\left[\sqrt{k}\mathbb{Z}\right]$ is the complex conjugate of a discretely gauged Wu Chern-Simons theory associated to the lattice $\sqrt{k}\mathbb{Z}$, with background field given by $[\check{C}]_k$, see Section \ref{SecDGWCS}.

We should justify why we identify the WCS action in \eqref{EqAnSDk2} with the gauged theory, rather than with the prequantum theory. Recall that \eqref{EqPartFuncGaugWCS} implies that their partition functions have the same phase. However, while the prequantum theory is invertible, the gauged theory is not if $|k| > 1$. It is known that self-dual fields with charges larger than 1 do not admit a single partition function, but rather a vector of conformal blocks \cite{Belov:2006jd}. This is the hallmark of theories with non-invertible anomaly field theories \cite{Monnierd}, and suggests the identification with the gauged theory rather than the prequantum theory.

We will see that the non-invertibility of $\overline{\textcal{WCS}}_{\rm G}\left[\sqrt{k}\mathbb{Z}\right]$ is also ultimately responsible for the appearance of the conformal blocks of the $A_{k-1}$ (2,0) SCFTs.

\subsection{Anomaly field theory of a free tensor multiplet}

The tensor multiplet is the (2,0) supersymmetric multiplet generated by the supercharges from the self-dual field theory. On a 6-dimensional $(2,0)$-manifold $M$, it contains symplectic Majorana fermions valued in the spinor bundle of $TM \oplus \mathscr{N}_M$, as well as anomaly-free scalars. We therefore obtain the anomaly of the charge $k$ tensor multiplet by adding the fermionic anomaly to the anomaly of the charge $k$ self-dual field. The former is the same as the fermionic anomaly on the worldvolume of a single M5-brane, already discussed in Section \ref{SecAnM5WV}. The anomaly of the charge $k$ tensor multiplet therefore reads
\be
\label{EqAnTMk3}
\frac{1}{2\pi i} \ln {\rm An}_{{\rm TM}(k)}(U,\check{C}_U) = -\frac{1}{2} \xi_f(U) -\frac{k-1}{8} \mathrm{\eta}_\sigma(U) -  k {\rm hs}(U,\omega)  - S_{WCS}(U,\sqrt{k}\mathbb{Z},\check{C}_U,\omega) \;,
\ee
The corresponding anomaly field theory is
\be
\label{EqAFTTMk}
\textcal{An}_{{\rm TM}(k)} = \left( \textcal{DF}^{\frac{1}{2}}_f \right)^{\otimes (-1)} \left( \textcal{DF}^{\frac{1}{4}}_\sigma \right)^{\otimes (-k+1)} \otimes \textcal{HS}^{\otimes (-k)} \otimes  \left( \overline{\textcal{WCS}}_{\rm G} [\sqrt{k}\mathbb{Z} \right)^{\otimes (-1)}
\ee
Compared to the anomaly \eqref{EqAFTSDk} of a charge $k$ self-dual field, there is an extra tensor product with the inverse of the half Dai-Freed theory $\textcal{DF}_f^{\frac{1}{2}}$. This field theory was already described in Section \ref{SecAFTM5WV}.

\section{Anomaly field theories of (2,0) SCFTs}

\label{Sec7dimform}

We combine here the results of the previous sections to describe the anomaly field theory of the $A_n$ (2,0) SCFT. We will see that the expressions involving the parameter $n$ can be naturally reexpressed in terms of Lie algebra data, yielding conjectural anomaly field theories for the $(2,0)$ SCFTs in the $D$ and $E$ series. These conjectures are automatically consistent with the exceptional isomorphisms of low rank algebras in the $A$, $D$ and $E$ series.

\subsection{The $A_n$ case}

The anomaly of the $A_n$ (2,0) SCFT is obtained \cite{Harvey:1998bx, Monnier:2014txa} from the anomaly of a stack of $k = n+1$ M5-branes by subtracting the anomaly of the center of mass degrees of freedom, which form a charge $k$ tensor multiplet. We need to lift this subtraction procedure to the level of anomaly field theories.

We tensor the anomaly field theory of the M5-brane worldvolumes \eqref{EqAFTTrivStackM5} with the anomaly field theory of the Hopf-Wess-Zumino terms \eqref{EqAFTHWZ} to find the anomaly field theory of a stack of $k$ M5-branes $\textcal{An}_{{\rm Stack},k}$. To subtract the center of mass, we tensor it with the field theory complex conjugate to the anomaly field theory of the charge $k$ tensor multiplet \eqref{EqAFTTMk}. The (invertible) Hopkins-Singer theories $\textcal{HS}$ appear in the tensor in complex conjugate pairs, and therefore cancel. Results of Section \ref{AppDGWCSAndLatticeDec} show that the gauged Wu Chern-Simons theories associated to the cubic lattice $\Lambda$ and to the 1-dimensional lattice $\sqrt{k}\mathbb{Z}$ combine into a gauged Wu Chern-Simons theory associated to the $A_n$ lattice, with vanishing background field. The anomaly field theory of the $A_n$ (2,0) SCFT therefore reads
\be
\label{EqAFTAnSCFT}
\textcal{An}_{A_n} = \left(\textcal{DF}^{\frac{1}{2}}_{f} \right)^{\otimes (-n)} \otimes \left( \textcal{DF}^{\frac{1}{4}}_\sigma \right)^{\otimes (-n)} \otimes \textcal{An}_{\rm HWZ} \otimes \overline{\textcal{WCS}}_{\rm G}[A_n, 0] \;,
\ee
\be
\label{EqAFTHWZ2}
\textcal{An}_{\rm HWZ} =  \left( \textcal{WCS}_{\rm P}[\mathbb{Z}, -2\check{b}]\right)^{\otimes \frac{n(n+1)}{2}} \otimes
\left(\textcal{BF}[-2\check{b}, \check{C}']\right)^{\otimes \frac{n(n+1)}{2}} \otimes 
\left( \textcal{CSp}_2[\check{b}] \right)^{\otimes \frac{(n+2)(n+1)n}{6}} \;.
\ee
The field theories appearing in these expressions have all been described in Sections \ref{SecAFTM5WV}, \ref{SecAFTHWZ} and \ref{SecAFTSDK}. We made explicit the fact that the background field vanishes in the notation for the $A_n$ discretely gauged WCS theory $\overline{\textcal{WCS}}_{\rm G}[A_n, 0]$. We can interpret the field theory factors above on the Coulomb branch of the $(2,0)$ SCFT. The first two factors in \eqref{EqAFTAnSCFT} are due to the free tensor multiplets. The third factor comes from the Hopf-Wess-Zumino terms already discussed above. The fourth factor is responsible for the conformal blocks of the $(2,0)$ SCFT, as we discuss in Section \ref{SecConfBlocks}.

\subsection{The general case}

The various $n$-dependent quantities appearing in \eqref{EqAFTAnSCFT} and \eqref{EqAFTHWZ2} have natural Lie algebra interpretations: $n$ is the rank $r_{\mathfrak{su}_{n+1}}$, $n(n+2)$ is the dimension $|\mathfrak{su}_{n+1}|$ and $n+1$ is the dual Coxeter number ${\rm h}_{\mathfrak{su}_{n+1}}$. This yields a natural conjecture for the anomaly field theory of any $(2,0)$ SCFT associated to a Lie algebra $\mathfrak{g}$ of type A, D or E:
\be
\label{EqAFTGenSCFT}
\textcal{An}_{\mathfrak{g}} = \left(\textcal{DF}^{\frac{1}{2}}_{f} \right)^{\otimes (-r_\mathfrak{g})} \otimes \left( \textcal{DF}^{\frac{1}{4}}_\sigma \right)^{\otimes (-r_\mathfrak{g})} \otimes \textcal{An}_{\rm HWZ} \otimes  \overline{\textcal{WCS}}_{\rm G}[\Lambda_\mathfrak{g}, 0] \;,
\ee
\be
\label{EqAFTHWZGen}
\textcal{An}_{\rm HWZ} =  \left( \textcal{WCS}_{\rm P}[\mathbb{Z}, -2\check{b}]\right)^{\otimes \frac{r_\mathfrak{g} {\rm h}_\mathfrak{g}}{2}} \otimes
\left(\textcal{BF}[-2\check{b}, \check{C}']\right)^{\otimes \frac{r_\mathfrak{g} {\rm h}_\mathfrak{g}}{2}} \otimes 
\left( \textcal{CSp}_2[\check{b}] \right)^{\otimes \frac{|\mathfrak{g}| {\rm h}_\mathfrak{g}}{6}} \;,
\ee
where $\Lambda_\mathfrak{g}$ is the root lattice of $\mathfrak{g}$. One can check explicitly that the exponents are all integers for every $\mathfrak{g}$ of type A, D or E. The relevant data is summarized in the table below:
\be
\begin{array}{l|ll}
& |\mathfrak{g}| & {\rm h}_\mathfrak{g} \\ \hline
A_n & n^2 + 2n & n + 1 \\
D_n & 2n^2 - n & 2n - 2 \\
E_6 & 78 & 12 \\
E_7 & 133 & 18 \\
E_8 & 248 & 30 \\
\end{array} \;.
\ee
As \eqref{EqAFTGenSCFT} and \eqref{EqAFTHWZGen} are expressed in term of Lie algebra data, they are automatically compatible with the exceptional isomorphisms between low rank algebras in the A, D and E series.

\subsection{Defects}

It is often interesting to consider (2,0) SCFTs in the presence of various defects. We do not have a complete picture of the relation between the defects of the (2,0) SCFTs and those of the anomaly field theory, but we describe here a correspondence between a class 2-dimensional defects in the (2,0) SCFT and a class of 3-dimensional defects of its anomaly field theory. We also present an analysis suggesting that the codimension 2 defects of the SCFT do not affect anomalies, and therefore are not visible in the anomaly field theory.

\paragraph{2-dimensional defects} The 6d SCFTs have instantonic strings, which are 2-dimensional defects charged under the self-dual fields. The self-duality condition requires them to carry both electric and magnetic charges. Their charges live in the weight lattice $\Lambda^\ast_\mathfrak{g}$ of the Lie algebra $\mathfrak{g}$. As the "W-bosons" of the 6d SCFT have charges in the root lattice $\Lambda_\mathfrak{g}$, a screening mechanism can neutralize the defect charges if they live in $\Lambda_\mathfrak{g}$. The observable charges therefore live in the finite discrete group $\Gamma_\mathfrak{g} = \Lambda_\mathfrak{g}^\ast/\Lambda_\mathfrak{g}$. The fact that these defects source the self-dual fields magnetically means that a 3-sphere linking the 2-dimensional worldvolume of the defect carries a flux of the self-dual field strength. Note that in order to have a well-defined configuration of the self-dual gauge field, we need to excise the worldsheet of the instantonic string.

To understand how these defects should be incorporated in the anomaly field theory, we use the fact that by definition, the 6d SCFT on a manifold $M$ can be used as a boundary condition of the anomaly field theory on a manifold $U$ with $\partial U = M$. We work on the Coulomb branch, where the gauge symmetry of the (2,0) SCFT is broken to the diagonal $U(1)^n$ subgroup. We model the self-dual gauge field, following \cite{Belov:2006jd}, as an ordinary gauge field $\check{B} = (b,B,H) \in \check{C}^3(M;\Lambda_\mathfrak{g})$ whose field strength $H \in \Omega^3(M)$ is required to live in a Lagrangian subgroup of $\Omega^3(M)$. The self-dual field strength of the self-dual field is then the self-dualization of $H$. The degree 3 gauge field $\check{A} = (a,A,F)$ of the Wu Chern-Simons theory in the anomaly field theory, which so far has been set to zero, should be trivialized by the self-dual field on the boundary: $\check{A} = d\check{B}$ on $M$, or more explicitly
\be
a = db \;, \quad A = H - dB - b \;, \quad F = dH \;.
\ee
Of course, if $\check{B}$ is a differential cocycle, then $\check{A} = 0$. However, the presence of an instantonic string forces $\check{B}$ to be a non-closed differential cochain, and therefore corresponds to configuration of the anomaly field theory where the Wu Chern-Simons gauge field is turned on. Indeed, although we do have $dH = 0$ and $db = 0$, the closedness condition $H - dB - b = 0$ is impossible to satisfy unless the fluxes of $H$ are valued in $\Lambda_\mathfrak{g}$. Therefore a $\Lambda^\ast_\mathfrak{g}$-valued flux of $H$ has to be accompanied with a non-zero $A$: the Wu Chern-Simons gauge field has a $\Gamma_\mathfrak{g}$-valued holonomy (the higher dimensional equivalent of a Wilson line) along the 3-spheres linking the worldsheet of the instanton string.

This also tells us that the worldsheet of the instanton string, that has already been excised from the boundary, has to be extended in $U$ as a 3-manifold with boundary and excised from $U$ as well. Without this operation, it would be impossible for $\check{A}$ to have holonomy along the linking 3-spheres.

In summary, the instantonic strings of the 6d SCFT correspond to discrete $\Lambda^\ast_\mathfrak{g}$-valued holonomies of the Wu Chern-Simons gauge field of the anomaly field theory along 3-spheres linking the worldsheet of the string. In the 7-dimensional spacetime of the anomaly field theory, the instantonic strings have to be promoted to 3-dimensional excised defects, including a holonomy of the Wu Chern-Simons gauge field along the linking 3-spheres.  

We can generalize the discussion above to arbitrary $\Lambda^\ast_\mathfrak{g}$-valued fluxes of the self-dual field. Such a flux corresponds to an element in $H^3(M;\Lambda^\ast_\mathfrak{g})$. $\Lambda^\ast_\mathfrak{g}$-valued fluxes generally require that $\check{B}$ is not closed, so $\check{A}$ cannot vanish. We therefore find again that the fluxes of $\check{B}$ are related to the holonomies of $\check{A}$. This relation is given by the surjective homomorphism $H^3(M;\Lambda^\ast_\mathfrak{g}) \rightarrow {\rm Hom}(H_3(M;\Lambda_\mathfrak{g}), U(1))$, induced by the evaluation of cocycles on cycles. The former group is the group of fluxes of $\check{B}$, while the latter group is the group of holonomies of $\check{A}$.

\paragraph{Codimension 2 defects} The most interesting class of defects of $(2,0)$ SCFTs are the codimension 2 defects. They are in particular crucial to the construction of many 4-dimensional supersymmetric quantum field theories from the $(2,0)$ SCFTs \cite{Witten:1997sc, Gaiotto:2009we, Gaiotto:2009hg}. In the M-theory realization of the $A_n$ SCFT, such defects are associated to M5-branes intersecting the stack of M5-branes along codimension 2 submanifolds.

These defects should correspond to codimension 2 defects of the anomaly field theory, but we have not been able to find natural candidates. The computation of the Chern-Simons term on the 4-sphere bundle $\tilde{W}$ over $W$, in the presence of defect M5-branes is strictly speaking ill-defined, because of the singularities of the C-field at the locus of the defect M5-branes. A simple counting argument reveal however that there should not be any cross term between the C-field field strength $G_{\rm stack}$ sourced by the stack and the field strength $G_{\rm defect}$ sourced by the defect M5-branes. Indeed, both forms have two legs along two of the three common transverse directions, showing that their wedge product vanishes. (Note that this argument is valid both in the physical 11-dimensional spacetime and in the 13-dimensional spacetime used to compute anomalies, because the codimensions are the same.)

This rough argument suggests that the inclusion of codimension 2 defects should not change the anomalies of the 6d SCFTs, except for restricting the group of allowed diffeomorphisms/R-symmetry transformations to the subgroup preserving the defects. This would explain why such defects seem invisible to the anomaly field theory. 

\section{The conformal blocks of the $(2,0)$ SCFTs}

\label{SecConfBlocks}

\paragraph{Dimension} An interesting feature of the anomaly field theory \eqref{EqAFTGenSCFT} is that it is not invertible \cite{Freed:2014iua}. In particular, its state space on a 6-dimensional $(2,0)$-manifold $M$ generally has dimension higher than 1. Indeed, all the tensor factors of \eqref{EqAFTGenSCFT} are invertible, except possibly for the discretely gauged Wu Chern-Simons theory $\overline{\textcal{WCS}}_{\rm G}[\Lambda_\mathfrak{g}, 0]$, whose state space has dimension $|H^3(M; \Gamma_\mathfrak{g})|^{1/2}$. $\Gamma_\mathfrak{g} = \Lambda_\mathfrak{g}^\ast/\Lambda_\mathfrak{g}$ is here the center of the simply connected group associated to $\mathfrak{g}$, and $|H^3(M; \Gamma_\mathfrak{g})|^{1/2}$ is an integer because $H^3(M; \Gamma_\mathfrak{g})$ carries a non-degenerate skew-symmetric pairing. The $(2,0)$ SCFT associated to the ADE Lie algebra $\mathfrak{g}$ has therefore a vector of partition functions taking value in a Hilbert space of dimension $|H^3(M; \Gamma_\mathfrak{g})|^{1/2}$. (The only $(2,0)$ SCFT that has an invertible anomaly field theory is the one associated to $E_8$. Indeed, $E_8$ has a unimodular root lattice, so $\Gamma_{E_8} = 1$ and the discretely gauged Wu Chern-Simons theory coincides with the prequantum theory, which is invertible.)

The dimension of the space of conformal blocks described above has been deduced previously from the reduction of the $(2,0)$ SCFT on a torus \cite{Witten:1998wy}, see also \cite{Henningson:2010rc, Tachikawa:2013hya}, as well as \cite{DelZotto:2015isa} for the case of (1,0) SCFTs. The proposed anomaly field theory is therefore consistent with the expected dimension of the conformal blocks of the $(2,0)$ theory.

\paragraph{A subtlety in the presence of torsion} In \cite{Witten:1998wy}, the space of conformal blocks was obtained by quantizing the action of discrete $H^3(M;\Gamma_\mathfrak{g})$-valued shifts of a background degree 3 $\Lambda_\mathfrak{g}$-valued gauge field minimally coupled to the self-dual field. It was claimed that upon quantization, the shift operators form a copy of the Heisenberg group $\mathsf{H}'$ associated to $H^3(M; \Gamma_\mathfrak{g})$ and its skew-symmetric pairing, and that moreover the action of $\mathsf{H}'$ on the space of conformal blocks is irreducible. As we will explain below, this is true only in the absence of torsion in $H^3(M; \Lambda^\ast_\mathfrak{g})$ (or equivalently in $H^3(M; \mathbb{Z})$). Indeed, in \cite{Witten:1998wy}, the irreducibility of the action was shown only in the absence of torsion.

A background degree 3 $\Lambda_\mathfrak{g}$-valued gauge field can be represented by a degree 4 $\Lambda_\mathfrak{g}$-valued differential cocycle $\check{C} = (c, C, G)$. The equivalence class of the gauge field is the associated differential cohomology class in $\check{H}^4(M;\Lambda_\mathfrak{g})$. We need to understand how $H^3(M; \Gamma_\mathfrak{g})$ acts on $\check{H}^4(M;\Lambda_\mathfrak{g})$: the action of the Heisenberg group $\mathsf{H}'$ on the space of conformal blocks should lift this action. For this, we can use the following fact proven in Proposition 3.1 of \cite{Monnier:2016jlo}, and already mentioned in Section \ref{SecDGWCS}. Classes in $H^3(M; \Gamma_\mathfrak{g})$ can be represented by differential cocycles of the form $(a, f, 0)$, with $a$ a degree 4 $\Lambda_\mathfrak{g}$-valued cocycle and $f$ a degree 3 $\Lambda^\ast_\mathfrak{g}$-valued cochain. The cocycle condition is $a = -df$, and two such cocycles are considered equivalent if $a_2 - a_1 = db$, $f_2 - f_1 = -b + dg$, for $b$ a degree 3 $\Lambda_\mathfrak{g}$-valued cochain and $g$ a degree 2 $\Lambda^\ast_\mathfrak{g}$-valued cochain. The equivalence classes of such cocycles is $H^3(M; \Gamma_\mathfrak{g})$. (This is due to the fact that $g$ is restricted to be a $\Lambda^\ast_\mathfrak{g}$-valued cochain instead of a $\Lambda_\mathfrak{g} \otimes_\mathbb{Z} \mathbb{R}$-valued cochain, as in the definition of standard differential cohomology.) A cocycle $(a,f,0)$ acts on $\check{C}$ by the obvious action $(c,C,G) \rightarrow (c + a, C + f, G)$, and one can show that this action descends to an action of $H^3(M; \Gamma_\mathfrak{g})$ on $\check{H}^4(M;\Lambda_\mathfrak{g})$.

In order to study the irreducibility of this action, we need to describe in some detail the structure of $H^3(M; \Gamma_\mathfrak{g})$ in the presence of torsion. Recall first the short exact sequence
\be
\label{EqShExSeqTors}
0 \rightarrow H^3_{\rm tors}(M; \Lambda^\ast_\mathfrak{g}) \rightarrow H^3(M; \Lambda^\ast_\mathfrak{g}) \rightarrow H^3_{\rm free}(M; \Lambda^\ast_\mathfrak{g}) \rightarrow 0 \;,
\ee
which holds for cohomology valued in any free abelian group. In addition, there is a long exact sequence derived from the short exact sequence of abelian groups $0 \rightarrow \Lambda_\mathfrak{g} \rightarrow \Lambda^\ast_\mathfrak{g} \rightarrow \Gamma_\mathfrak{g} \rightarrow 0$, reading:
\be
... \rightarrow H^3(M; \Lambda_\mathfrak{g}) \stackrel{\iota}{\rightarrow} H^3(M; \Lambda^\ast_\mathfrak{g}) \rightarrow H^3(M; \Gamma_\mathfrak{g}) \rightarrow H^4(M; \Lambda_\mathfrak{g}) \stackrel{\iota}{\rightarrow} H^4(M; \Lambda^\ast_\mathfrak{g}) \rightarrow ...
\ee
that tells us that the $\Gamma_\mathfrak{g}$-valued cohomology fits in a short exact sequence
\be
0 \rightarrow H^3(M; \Lambda^\ast_\mathfrak{g})/\iota(H^3(M; \Lambda_\mathfrak{g})) \rightarrow H^3(M; \Gamma_\mathfrak{g}) \rightarrow {\rm ker}\left ( \iota |_{H^4(M; \Lambda_\mathfrak{g})} \right) \rightarrow 0 \;.
\ee
Combining it with \eqref{EqShExSeqTors}, we obtain the following filtration for $H^3(M; \Gamma_\mathfrak{g})$:
\be
\label{EqFiltH3Gamg}
\mathsf{T} \hookrightarrow \mathsf{C} \hookrightarrow H^3(M; \Gamma_\mathfrak{g}) \;,
\ee
\be
\mathsf{T} := H^3_{\rm tors}(M; \Lambda^\ast_\mathfrak{g}) /\iota(H^3(M; \Lambda_\mathfrak{g})) \;, \quad \mathsf{C} := H^3(M; \Lambda^\ast_\mathfrak{g}) /\iota(H^3(M; \Lambda_\mathfrak{g})) \;,
\ee
\be
\mathsf{G} := \mathsf{C}/\mathsf{T} \simeq H^3_{\rm free}(M; \Lambda^\ast_\mathfrak{g}) /\iota(H^3(M; \Lambda_\mathfrak{g})) \;,
\ee
\be
\mathsf{K} := H^3(M; \Gamma_\mathfrak{g})/\mathsf{C} \simeq {\rm ker}\left ( \iota |_{H^4(M; \Lambda_\mathfrak{g})} \right) \;.
\ee
$\mathsf{G}$ can alternatively be described as the space of de Rham cohomology classes with periods valued in $\Lambda^\ast_\mathfrak{g}$, modulo the space of de Rham cohomology classes with periods valued in $\Lambda_\mathfrak{g}$. As the de Rham cohomology of degree 3 on $M$ carries a non-degenerate skew-symmetric pairing, the same is true for $\mathsf{G}$. The pairing coincides with the non-degenerate skew-symmetric pairing $B$ of $H^3(M; \Gamma_\mathfrak{g})$, restricted on $\mathsf{C}$ and induced on the quotient $\mathsf{C}/\mathsf{T}$. The fact that $B$ is non-degenerate also implies that $\mathsf{T}$ and $\mathsf{K}$ are Pontryagin duals of each other.

With these technical details cleared up, we can immediately see that the subgroup $\mathsf{T} \subset H^3(M,\Gamma_\mathfrak{g})$ acts trivially on $\check{H}^4(M,\Lambda_\mathfrak{g})$. Indeed, an element of $\mathsf{T}$ can be represented by a differential cocycle of the form $(0,f,0)$, with $f$ a $\Lambda^\ast_\mathfrak{g}$-valued cocycle. The fact that this cocycle comes from the torsion subgroup $H^3_{\rm tors}(M;\Lambda^\ast_\mathfrak{G})$ means that $f$, although not necessarily trivial, is trivial as a $\Lambda_\mathfrak{g} \otimes_\mathbb{Z} \mathbb{R}$-valued cocycle. Its action sends $\check{C} = (c,C,G)$ to $(c, C + f, G)$, an equivalent cocycle. 

We deduce that in the presence of torsion in $H^3(M;\mathbb{Z})$, the action of $H^3(M,\Gamma_\mathfrak{g})$ on $\check{H}^4(M,\Lambda_\mathfrak{g})$ has a kernel, and therefore the conformal blocks cannot form an irreducible representation of the Heisenberg group $\mathsf{H}'$, unlike what is claimed in \cite{Witten:1998wy}. We can explain this in slightly more physical terms as follows. The elements of $\mathsf{T}$ are dual to $\Lambda_\mathfrak{g}$-valued homology cycles of degree 2, so we could think of them as trying to shift the holonomies of the gauge field along those cycles. But a degree 3 gauge field has no holonomy along degree 2 cycles, so this action has to be trivial.

The correct picture can be understood by a careful construction \cite{Monnier:2016jlo} of the state space of the discretely gauged Wu Chern-Simons theory $\overline{\textcal{WCS}}_{\rm G}[\Lambda_\mathfrak{g}, 0]$. As $\mathsf{T}$ acts trivially, the action passes to an action of $H^3(M;\Gamma_\mathfrak{g})/\mathsf{T}$, which contains $\mathsf{G}$ as a subgroup. As mentioned above, $\mathsf{G}$ carries a non-degenerate skew-symmetric pairing, with an associated Heisenberg group $\mathsf{H}$. The state space of the anomaly field theory is a direct sum of $|\mathsf{K}|$ irreducible representations of $\mathsf{H}$, of dimension $|\mathsf{K}| |\mathsf{G}|^{1/2}$. As the filtration \eqref{EqFiltH3Gamg} ensures that 
\be
|H^3(M;\Gamma_\mathfrak{g})| = |\mathsf{T}| |\mathsf{G}| |\mathsf{K}|
\ee
and the perfect pairing between $\mathsf{T}$ and $\mathsf{K}$ implies that $|\mathsf{T}| = |\mathsf{K}|$, the dimension of the state space is equal to $|H^3(M;\Gamma_\mathfrak{g})|^{1/2}$, even in the presence of torsion.


\subsection*{Acknowledgments}

This research has been supported in part by SNSF grants No. 152812, 165666, and by NCCR SwissMAP, funded by the Swiss National Science Foundation.

\appendix

\section{The differential cohomology model of abelian gauge fields}

\label{AppDiffCohom}

In this appendix, we recall how differential cocycles and differential cohomology \cite{springerlink:10.1007/BFb0075216, hopkins-2005-70} can be used to model higher abelian gauge fields and their gauge equivalence classes. A pedagogical and physically motivated reference for gauge group $U(1)$ appears in Section 2 of \cite{Freed:2006yc}. We take a more general view in the present appendix, including arbitrary compact connected abelian gauge groups, as well as gauge fields with shifted fractional flux quantization conditions. 

\paragraph{Shifted differential cochains} Let $\Lambda$ be an integral lattice such that the abelian gauge group takes the form $V/\Lambda$, where $V = \Lambda \otimes_\mathbb{Z} \mathbb{R}$. We write $C^p(M;A)$ for the group of degree $p$ cochains with value in an abelian group $A$ on a smooth manifold $M$, and $Z^p(M;A)$ for the corresponding group of cocycles. Choose a cocycle $s \in Z^p(M;V/\Lambda)$. Let $\Omega^p(M;V)$ be the group of $V$-valued degree $p$ smooth differential forms on $M$. The set of \emph{degree $p$ $\Lambda$-valued differential cochains on $M$ shifted by s} is
\be
\check{C}^p_s(M; \Lambda) := \left\{ (g,C,G) \in C^p(M;V) \times C^{p-1}(M;V) \times \Omega^p(M;V) | g = s \mbox{ mod } \Lambda \right\} \;.
\ee
$\check{C}^p(M; \Lambda) = \check{C}^p_0(M; \Lambda)$ is an abelian group (with the group structure induced by the addition of cochains and forms) and $\check{C}^p_s(M; \Lambda)$ is a torsor over $\check{C}^p(M; \Lambda)$.

We write differential cochains with a caron: $\check{C} = (g,C,G)$. $[\check{C}]_{\rm ch} := g$ is the \emph{characteristic}, $[\check{C}]_{\rm co} := C$ is the \emph{connection} and $[\check{C}]_{\rm cu} := G$ is the \emph{curvature} or \emph{field strength} of $\check{C}$. Differential cochains $\check{C}$ with $[\check{C}]_{\rm ch} = 0$ are called \emph{topologically trivial}, while differential cocycles with $[\check{C}]_{\rm cu} = 0$ are called \emph{flat}

\paragraph{Shifted differential cocycles} We define a differential on the complex $\check{C}^\bullet_s(M;\Lambda)$ as follows:
\be
d(g,C,G) := (dg, G - g - dC, dG) \;,
\ee
which satisfies $d^2 = 0$. The set of \emph{degree $p$ $\Lambda$-valued differential cocycles on $M$ shifted by $s$} $\check{Z}_s^p(M;\Lambda)$ is the kernel of $d$ on $\check{C}^p_s(M; \Lambda)$. $\check{Z}^p(M;\Lambda) = \check{Z}_0^p(M;\Lambda)$ is an abelian group and $\check{Z}_s^p(M;\Lambda)$ is a torsor over $\check{Z}_0^p(M;\Lambda)$.

We can put an equivalence relation on the set of degree $p$ differential cocycles, by seeing as equivalent any pair of differential cocycles differing by the differential of a flat unshifted differential cochain:
\be
\label{EqEquivDiffCoc}
(g,C,G) \sim (g + dh, C - h - dB, G)
\ee
for all $(h,B,0) \in \check{C}_0^{p-1}(M;\Lambda)$.

The \emph{degree $p$ $\Lambda$-valued differential cohomology group of $M$ shifted by $s$} $\check{H}^p_s(M;\Lambda)$ is the quotient of $(\check{Z}^\bullet_s(M;\Lambda),d)$ by this equivalence relation. We write $\check{H}^p(M;\Lambda) := \check{H}^p_0(M;\Lambda)$.

\paragraph{Physical interpretation} $\check{Z}_s^p(M;\Lambda)$ models degree $p-1$ abelian gauge fields on $M$ for the gauge group $V/\Lambda$, with a shifted flux quantization condition determined by $s$. The gauge transformations correspond to the equivalences \eqref{EqEquivDiffCoc}, and the elements of $\check{H}_s^p(M;\Lambda)$ are gauge equivalence classes of fields.

In the physics literature, it is common to model gauge fields as differential forms. This is possible only if the gauge field has trivial topology. For ordinary degree 1 gauge fields, this occurs when the gauge field is a connection on a trivial principal bundle. Topologically trivial differential cocycles are triplets $(0, C, G)$ satisfying $G = dC$, where the $V$-valued cochain $C$ can be seen as a degree $p-1$ $V$-valued differential form. The gauge transformations preserving $g = 0$ are $C \rightarrow C - dB$ for $B$ a degree $p-2$ $V$-valued cochain that can as well be seen as a degree $p-2$ differential form. $G$ can therefore be interpreted as the field strength of the gauge field, and $C$ as the gauge field itself. The advantage of using differential cocycles is that gauge fields of arbitrary topology can be modelled. 

The transformations \eqref{EqEquivDiffCoc} with $h \neq 0$ correspond to large gauge transformations. The holonomy ("Wilson line") of the gauge field along a $p-1$-cycle $Z$ is given by $\exp 2 \pi i \int_Z C$, which is checked to be gauge invariant.

The constraint $G - g - dC = 0$ shows that the periods of $G$ coincide with the periods of $g$, and will therefore be given mod $\Lambda$ by the cocycle $s$. Ordinary gauge fields, whose fluxes are valued in $\Lambda$, correspond therefore to unshifted differential cocycles, associated to $s = 0$. Shifted differential cocycles model gauge fields with shifted fractional quantization law. 

For instance, the M-theory C-field is a $U(1)$-valued degree 3 gauge field, whose fluxes are given mod 1 by half the periods of the fourth Stiefel-Whitney class $w_4$ of spacetime \cite{Witten:1996md}. It is naturally seen as an element of $\check{Z}^4_{w_4}(M;\mathbb{Z})$, where we see $w_4$ as a $\mathbb{R}/\mathbb{Z}$-valued cocycle using the standard embedding $\mathbb{Z}_2 \subset \mathbb{R}/\mathbb{Z}$.

\paragraph{Cup product} \cite{springerlink:10.1007/BFb0075216, hopkins-2005-70} The pairing on $\Lambda$ induces a product on the groups of unshifted differential cochains
\begin{align}
\label{EqDefCupProdDiffCohom}
\cup: \: & \check{C}^p_0(M; \Lambda) \times \check{C}^q_0(M; \Lambda) \rightarrow \check{C}^{p+q}_0(M; \mathbb{Z}) \\
& \check{C}_1 \cup \check{C}_2 = (g_1 \cup g_2, (-1)^{p} g_1 \cup C_2 + C_1 \cup G_2 + H^\wedge_\cup(G_1, G_2) , G_1 \wedge G_2) \notag
\end{align}
where we wrote $\check{C}_i = (g_i, C_i, G_i)$. $H^\wedge_\cup$ is a choice of homotopy between the wedge and cup products, i.e. a homomorphism from $\Omega^p(M; V) \times \Omega^q(M; V)$ to $C^{p+q-1}(M; V)$ such that
\be
dH^\wedge_\cup(G_1, G_2) + H^\wedge_\cup(dG_1, G_2) + (-1)^{p} H^\wedge_\cup(G_1, dG_2) = G_1 \wedge G_2 - G_1 \cup G_2 \;.
\ee
$H^\wedge_\cup$ can be chosen canonically if a suitable model for cochains is used, see \cite{springerlink:10.1007/BFb0075216}. One can check that $\cup$ passes to a well-defined product on differential cocycles and differential cohomology classes.

There is no obvious way of defining a cup product on shifted differential cochains for general shifts.

\section{Wu structures}

\label{AppWuStruct}

A more detailed account of Wu structures can be found in Appendix C of \cite{Monnier:2016jlo}.

Wu structures are higher analogues of spin structures. To understand their definition, it is useful to recall the definition of a spin structure. Let $BSO(n)$ be the classifying space of bundles with $SO(n)$ structure. To any smooth oriented manifold $M$ of dimension $n$, we can associate its tangent bundle $TM$, and therefore a (homotopy class of) classifying map from $M$ into $BSO(n)$. The second Stiefel-Whitney class $w_2$ can be seen as a homotopy class of maps from $BSO(n)$ into $K(\mathbb{Z}_2, 2)$. The associated homotopy fiber is written $BSpin(n)$, and a \emph{spin structure} on $M$ is a lift of the classifying map of $TM$ from $BSO(n)$ to $BSpin(n)$. 

Wu classes form a family of $\mathbb{Z}_2$-valued characteristic classes that can be expressed in terms of the Stiefel-Whitney classes, see for instance \cite{opac-b1077949} Chapter 11 for a definition. For oriented manifolds, the second Stiefel-Whitney class coincides with the second Wu class. A Wu structure of degree $p$ is defined as above, by replacing the second Stiefel-Whitney class by the degree $p$ Wu class. Explicitly, if $\nu_p$ is the degree $p$ Wu class, we define $BSO[\nu_p](n)$ to be the homotopy fiber of the map from $BSO(n)$ into $K(\mathbb{Z}_2,p)$ defined by $\nu_p$. A \emph{Wu structure} on $M$ is a lift of the classifying map of $TM$ from $BSO(n)$ to $BSO[\nu_p](n)$. As is obvious from the above discussion, a Wu structure of degree 2 is nothing but a spin structure. In the present paper, we are mostly interested in Wu structures of degree $4$, associated to the Wu class of degree 4
\be
\nu_4(TM) = w_4(TM) + \left( w_2(TM) \right)^2 \;.
\ee

Wu structures of degree $p$, when they exist, are in bijection with $H^{p-1}(M; \mathbb{Z}_2)$. Any manifold of dimension strictly smaller than $2p$ admits Wu structures of degree $p$.

The Wu structure of $M$ is encoded in the homotopy class of the classifying map from $M$ into $BSO[\nu_p](n)$. In order to have a more concrete object representing the Wu structure, we can pick an actual classifying map and proceed as follows. Pick a cocycle representative of the Wu class on $BSO(n)$, which we also write $\nu_p$ for simplicity. Pull it back to $BSO[\nu_p](n)$. By definition, $\nu_p$ is exact on $BSO[\nu_p](n)$, so let us choose a trivialization $\eta$ on $BSO[\nu_p](n)$: $d\eta = \nu_p$. We can pull-back $\eta$ by the classifying map to obtain a cochain on $M$. This cochain encodes the data of the Wu structure on $M$. 

In the main text, we are only interested in degree 4 Wu structures, i.e. $p = 4$. We write $\nu_M$ for the degree 4 Wu cocycle on a manifold $M$.

\section{Euler structures}

\label{AppEulerStruct}

The same idea can be applied to the Euler class of an arbitrary real bundle $\mathscr{N}_M$ over $M$, defining an Euler structure on $\mathscr{N}_M$. Euler structures on the tangent space of 3-manifolds have been discussed previously in \cite{Turaev1997}. We make here the construction explicit in the case of interest to us, where $\mathscr{N}_M$ is a rank 5 bundle. $\mathscr{N}_M$ is classified by a homotopy class of maps into $BSO(5)$. The Euler class $e$ defines a homotopy class of maps from $BSO(5)$ into $K(\mathbb{Z}, 5)$, and we write $BSO[e](5)$ for the corresponding homotopy fiber. $BSO[e](5)$ carries a universal bundle $\mathscr{N}$, whose Euler class vanishes by definition. Let $\widetilde{BSO}[e](5)$ be the associated 4-sphere bundle. It is possible to pick a degree 4 integral cocycle $a$ on $\widetilde{BSO}[e](5)$ restricting to a generator of the top cohomology on each fiber. Writing $\pi_\ast$ for the push forward map over the fibers of $\mathscr{N}$, the discussion around (5.20) in \cite{Witten:1999vg} shows that
\be
\label{EqDefbIntGlobAngForm2}
b := \pi_\ast(a \cup a)
\ee
represents $w_4(\mathscr{N})$ when reduced modulo 2.

If the Euler class of $\mathscr{N}_M$ vanishes, an \emph{Euler structure} on $\mathscr{N}_M$ is a (homotopy class of) lift of its classifying map from $BSO(5)$ into $BSO[e](5)$. Denoting by $\tilde{M}$ the 4-sphere bundle of $\mathscr{N}_M$, we can pull back $a$ through a classifying map to obtain an integral cocycle $a_{\tilde{M}}$ restricting to a generator of the top cohomology on each fiber.

Like in the case of Wu structures, we will assume that the Euler structure includes a choice of classifying map to $BSO[e](5)$, rather than just a homotopy class of maps.

\section{The cobordism group of $(2,0)$-manifolds}

\label{SecCobGroup20Struct}

\subsection{Statement of the theorem}

Recall the definition of $(2,0)$ structures and morphisms of $(2,0)$-manifolds in Section \ref{SecPrelim}. In this appendix, we prove
\begin{theorem}
\label{Th7dim20ManBound}
Any 7-dimensional manifold $U$ endowed with a $(2,0)$-structure is the boundary of an 8-dimensional manifold $W$ endowed with a $(2,0)$-structure that restricts to the one of $U$ on the boundary, provided
\be
\label{EqCondBoundCob}
w_2(\mathscr{N}_U) w_3(\mathscr{N}_U) = 0 \;.
\ee
\end{theorem}
Condition \ref{EqCondBoundCob} is sufficient, but we do not know whether it is necessary or not. For reasons explained in Appendix C of \cite{Monnierb}, a $(2,0)$-manifold $U$ is the boundary of a $(2,0)$-manifold $W$ if and only if $U$ corresponds to a trivial class in the stable homotopy group
\be
\Omega^{M5}_{12} = \lim_{n \rightarrow \infty} \pi_{12+n}(MSpin(n) \wedge TF \wedge K(\mathbbm{Z},4)_+) = \tilde{\Omega}^{\rm Spin}_{12}(TF \wedge K(\mathbbm{Z},4)_+) \;.
\ee
In this formula, $F := BSO[e](5)$ is the homotopy fiber of the map 
\be
BSO(5) \stackrel{e}{\rightarrow} K(\mathbbm{Z},5)
\ee
defined by the Euler class of the universal bundle, $TF$ is the Thom space of the universal bundle on $F$ pulled-back from $BSO(5)$, $MSpin(n)$ is the Thom space of the universal bundle over $BSpin(n)$, $K(\mathbbm{Z},4)_+$ is the Eilenberg-MacLane space $K(\mathbbm{Z},4)$ with a disjoint base point and $\tilde{\Omega}^{\rm Spin}_{\bullet}$ are the reduced spin cobordism groups. As also explained in \cite{Monnierb}, we can rewrite $\Omega^{M5}_{12}$ as follows:
\be
\label{EqbordGroup20Struct}
\Omega^{M5}_{12} = \tilde{\Omega}^{\rm Spin}_{12}(TF) \oplus \tilde{\Omega}^{\rm Spin}_{12}(TF \wedge K(\mathbbm{Z},4)) \;.
\ee
In the following, we will use the Atiyah-Hirzebruch spectral sequence to show that the second group on the right-hand side vanishes and that the first one is either zero or $\mathbb{Z}_2$. The potential obstruction is generated by the (homology dual of the) characteristic class $w_2(\mathscr{N})w_3(\mathscr{N})$, so if the latter vanishes, the 7-dimensional $(2,0)$-manifold bounds.

\subsection{The integral homology of $F$}

Our first task is to compute the low degree homology groups of $F$ with coefficients in $\mathbb{Z}$ and $\mathbb{Z}_2$. The fibration $F \rightarrow BSO(5) \rightarrow K(\mathbb{Z},5)$ implies the existence of a fibration
\be
K(\mathbb{Z},4) \rightarrow F \rightarrow BSO(5) \;.
\ee
We use the Serre spectral sequence of this fibration to compute the cohomology of $F$ and then use the universal coefficient theorem to deduce the homology.

Low degree homology groups of $K(\mathbb{Z},4)$ with integral coefficients are given for instance the Appendix C of \cite{Breen2016}. We use the universal coefficient theorem to deduce from them the low degree cohomology with integral coefficients
\be
\label{EqCohomKZ4Z}
H^\bullet(K(\mathbb{Z},4);\mathbb{Z}) = \left( \begin{array}{cccccccccc}
0 & 1 & 2 & 3 & 4 & 5 & 6 & 7 & 8 & 9\\
\mathbb{Z} & 0 & 0 & 0 & \mathbb{Z} & 0 & 0 &  \mathbb{Z}_2 & \mathbb{Z} & \mathbb{Z}_3 \\
1 & - & - & - & g & - & - & \beta \circ {\rm Sq}^2(g) & g^2 & ?
\end{array} ... \right) \;,
\ee
where we wrote the generators in terms of the universal class $g \in H^4(K(\mathbb{Z},4);\mathbb{Z})$. $\beta$ is the Bockstein of the short exact sequence $\mathbb{Z} \rightarrow \mathbb{Z} \rightarrow \mathbb{Z}_2$ and $Sq^2$ is the second Steenrod square, implicitly precomposed with reduction mod 2.

The cohomology of $BSO(n)$ is described in \cite{Brown1982}. We have
\be
\begin{aligned}
\label{EqCohomBSO5Z}
H^\bullet &(BSO(5);\mathbb{Z}) = \\
&\left( \begin{array}{cccccccccc}
0 & 1 & 2 & 3 & 4 & 5 & 6 & 7 & 8 & 9\\
\mathbb{Z} & 0 & 0 & \mathbb{Z}_2 & \mathbb{Z} & \mathbb{Z}_2 & \mathbb{Z}_2 & \mathbb{Z}_2^2 & \mathbb{Z}^2 \oplus \mathbb{Z}_2 & \mathbb{Z}_2^2\\
1 & - & - & W_2 & p_1 & W_4 & W_2^2 & p_1 W_2, W_{2,4} & p_2, p_1^2, W_2 W_4 & W_2^3, p_1 W_4
\end{array} ... \right) \;,
\end{aligned}
\ee
where $\{p_i\}$ are the Pontryagin classes and $\{W_i\}$ are the integral Stiefel-Whitney classes, defined from the Stiefel-Whitney classes $\{w_i\}$ by $W_i = \beta w_i$. We also wrote $W_{2,4} = \beta(w_2 w_4)$. The Euler class $e$ coincides with $W_4$. 

There is a Serre spectral sequence
\be
E^2_{p,q} = H^p(BSO(5), H^q(K(\mathbb{Z},4);\mathbb{Z})) \Rightarrow H^{p+q}(F; \mathbb{Z}) \;.
\ee
whose second page has the following form.
\be
\label{EqSecPagSSSF}
\begin{array}{c|cccccccccc}
9 & \mathbb{Z}_3 & 0 & ... & ... & ... & ... & ... & ... & ... & ...\\
8 & \mathbb{Z} & 0 & 0 & ... & ... & ... & ... & ... & ... & ...\\
7 & \mathbb{Z}_2 & 0 & \mathbb{Z}_2 & \mathbb{Z}_2 & ... & ... & ... & ... & ... & ...\\
6 & 0 & 0 & 0 & 0 & 0 & 0 & 0 & 0 & 0 & 0\\
5 & 0 & 0 & 0 & 0 & 0 & 0 & 0 & 0 & 0 & 0\\
4 & \mathbb{Z} & 0 & 0 & \mathbb{Z}_2 & \mathbb{Z} & \mathbb{Z}_2 & \mathbb{Z}_2 & \mathbb{Z}_2^2 & \mathbb{Z}^2 \oplus \mathbb{Z}_2 & \mathbb{Z}_2^2\\
3 & 0 & 0 & 0 & 0 & 0 & 0 & 0 & 0 & 0 & 0\\
2 & 0 & 0 & 0 & 0 & 0 & 0 & 0 & 0 & 0 & 0\\
1 & 0 & 0 & 0 & 0 & 0 & 0 & 0 & 0 & 0 & 0\\
0 & \mathbb{Z} & 0 & 0 & \mathbb{Z}_2 & \mathbb{Z} & \mathbb{Z}_2 & \mathbb{Z}_2 & \mathbb{Z}_2^2 & \mathbb{Z}^2 \oplus \mathbb{Z}_2 & \mathbb{Z}_2^2\\
\hline q/p  & 0 & 1 & 2 & 3 & 4 & 5 & 6 & 7 & 8 & 9
\end{array}
\ee
By the definition of $F$, the Euler class pulls back to a trivial class. The only way that this can occur in the spectral sequence above is if the 5th differential satisfies $d^5_{0,4}(g) = W_4$. This determines the cohomology groups of $F$ in degrees 0 to 6 to be $\mathbb{Z}$, 0, 0, $\mathbb{Z}_2$, $\mathbb{Z}^2$, 0, $\mathbb{Z}_2$. The extra free generator in degree $4$ compared to $BSO(5)$ is $2g$, coming from the $2E^2_{0,4} = E^\infty_{0,4} = 2\mathbb{Z}$ term of the spectral sequence.

We will need the integral homology of $F$ in degree 7, so we have to determine the integral cohomology of $F$ in degree 7 as well as the torsion part of the cohomology in degree 8.

\paragraph{Degree 7} Clearly, $H^7(F; \mathbb{Z})$ can only be pure torsion. $E^2_{7,0} = \mathbb{Z}_2^2$ survives through the spectral sequence. $E^2_{3,4}$ is generated by $gW_2$ and using the compatibility of the differentials with the cup product, we have $d^5(g W_2) = W_4 W_2 \neq 0$, so it does not contribute to the cohomology of $F$. $E^2_{0,7} = \mathbb{Z}_2$ is generated by $\beta \circ {\rm Sq}^2(g)$, which survives through the spectral sequence. 

\paragraph{Degree 8} $E^2_{8,0} = \mathbb{Z}^2 \oplus \mathbb{Z}_2$, with the torsion term generated by $W_4 W_2$. As we saw above, the latter is in the image of $d^5$, so is killed by the spectral sequence. $E^2_{4,4}$ is generated by $gp_1$ and $d^5(gp_1) = W_4 p_1 \neq 0$, so only the even multiples survive. $E^2_{0,8}$ is sent by $d^2$, $d^5$ and $d^9$ onto torsion groups. This cannot generate any torsion. We have therefore shown that $H^8(F; \mathbb{Z})$ has no torsion.

\paragraph{Result} The discussion above and the universal coefficient theorem yield the integral homology groups of $F$:
\be
\label{EqHomFIntCoeff}
\begin{aligned}
H_\bullet & (F;\mathbb{Z}) = \\
& \left( \begin{array}{cccccccccc}
0 & 1 & 2 & 3 & 4 & 5 & 6 & 7\\
\mathbb{Z} & 0 & \mathbb{Z}_2 & 0 & \mathbb{Z}^2 & \mathbb{Z}_2 & \mathbb{Z}_2^3 &  0\\
1 & - & (W_2)^\ast & - & p_1^\ast, (2g)^\ast & (W_2^2)^\ast & (p_1 W_2)^\ast, (W_{2,4})^\ast, (\beta {\rm Sq}^2(g))^\ast & -
\end{array} ... \right)
\end{aligned}
\ee
A cohomology class decorated with an asterisk denotes the dual homology class with respect to the basis of generators chosen in \eqref{EqCohomKZ4Z} and \eqref{EqCohomBSO5Z}. The 7th homology group vanishes because the 7th cohomology group has no free part and the 8th cohomology group has no torsion.

\subsection{The homology of $F$ with $\mathbb{Z}_2$-coefficients}

The cohomology of $BSO(5)$ with $\mathbb{Z}_2$ coefficient can also be found in \cite{Brown1982}. It is expressed in terms of the Stiefel-Whitney classes as
\be
\label{EqCohomBSO5Z2}
\begin{aligned}
H^\bullet & (BSO(5);\mathbb{Z}_2) = \\
&\left( \begin{array}{cccccccc}
0 & 1 & 2 & 3 & 4 & 5 & 6 & 7\\
1 & - & w_2 & w_3 & w_4, w_2^2 & w_5, w_3 w_2 & w_4 w_2, w_3^2, w_2^3 & w_5 w_2, w_3 w_2^2, w_4 w_3
\end{array} ... \right) \;,
\end{aligned}
\ee
where each generator generates a $\mathbb{Z}_2$ subgroup.

To compute the the homology of $F$ with $\mathbb{Z}_2$ coefficients we can repeat our analysis of the Serre spectral sequence above for $\mathbb{Z}_2$ coefficients. Alternatively, the universal coefficient theorem allows us to deduce it from \eqref{EqHomFIntCoeff}. A comparison with the cohomology of $BSO(5)$ above allows to identify the generators as follows:
\be
\label{EqHomFZ2Coeff}
\begin{aligned}
 & H_\bullet (F;\mathbb{Z}_2) = \\
&\left( \begin{array}{cccccccccc}
0 & 1 & 2 & 3 & 4 & 5 & 6\\
1 & - & (w_2)^\ast & (w_3)^\ast & (w_4)^\ast, (w_2^2)^\ast & (w_3 w_2)^\ast & (w_4 w_2)^\ast, (w_3^2)^\ast, (w_2^3)^\ast, ({\rm Sq}^2 h)^\ast
\end{array} ... \right)
\end{aligned}
\ee
given in terms of the homology basis dual to \eqref{EqCohomBSO5Z2}. We wrote $h$ for the generator of $H^4(K(\mathbb{Z},4);\mathbb{Z}_2)$. We see that the mod 2 reduction of the Euler class, which coincides with $w_5$, is killed.

\subsection{The 12th spin cobordism group of $TF$}

To compute $\tilde{\Omega}^{\rm Spin}(TF)$, we use the Atiyah-Hirzebruch spectral sequence
\be
E^2_{p,q} = \tilde{H}_p(TF, \Omega^{\rm spin}_q({\rm pt.})) \Rightarrow \tilde{\Omega}^{\rm Spin}_{p+q}(TF) \;.
\ee
The spin cobordism groups of the point can for instance be found in Stong's appendix in \cite{Witten:1985bt}
\be
\Omega^{\rm spin}_\bullet({\rm pt.}) = \left( \begin{array}{cccccccc}
0 & 1 & 2 & 3 & 4 & 5 & 6 & 7 \\
\mathbb{Z} & \mathbb{Z}_2 & \mathbb{Z}_2 & 0 & \mathbb{Z} & 0 & 0 & 0
\end{array} ... \right) \;.
\ee
We will make use of the fact the second differential of the spectral sequence above coincides at $q = 0$ and $q = 1$ with the dual of the second Steenrod square composed with reduction mod 2 and with the dual of the second Steenrod square, respectively \cite{Zhubr2001}:
\be
d^2_{p,0} = ({\rm Sq}^2)^\ast \circ \rho_2 \;, \quad d^2_{p,1} = ({\rm Sq}^2)^\ast \;.
\ee
We can equivalently write $(d^2_{p,0})^\ast = \epsilon \circ {\rm Sq}^2$, where $\epsilon$ is the natural homomorphism $H^p(TF;\mathbb{Z}_2) \rightarrow {\rm Hom}(H_p(TF,\mathbb{Z}), \mathbb{Z}_2)$ given by the evaluation of representing cocycles on representing cycles.

The second page of the Atiyah-Hirzebruch spectral sequence is as follows:
\be
\label{EqSecPagAHSS}
\begin{array}{c|cccccccc}
7 & 0 & 0 & 0 & 0 & 0 & 0 & 0 & 0\\
6 & 0 & 0 & 0 & 0 & 0 & 0 & 0 & 0\\
5 & 0 & 0 & 0 & 0 & 0 & 0 & 0 & 0\\
4 & \mathbb{Z} & 0 & \mathbb{Z}_2 & 0 & \mathbb{Z}^2 & \mathbb{Z}_2 & \mathbb{Z}_2^3 &  0\\
3 & 0 & 0 & 0 & 0 & 0 & 0 & 0 & 0\\
2 & \mathbb{Z}_2 & 0 & \mathbb{Z}_2 & \mathbb{Z}_2 & \mathbb{Z}_2^2 & \mathbb{Z}_2 & \mathbb{Z}_2^4 & ... \\
1 & \mathbb{Z}_2 & 0 & \mathbb{Z}_2 & \mathbb{Z}_2 & \mathbb{Z}_2^2 & \mathbb{Z}_2 & \mathbb{Z}_2^4 & ... \\
0 & \mathbb{Z} & 0 & \mathbb{Z}_2 & 0 & \mathbb{Z}^2 & \mathbb{Z}_2 & \mathbb{Z}_2^3 &  0\\
\hline q/p  & 5 & 6 & 7 & 8 & 9 & 10 & 11 & 12
\end{array}
\ee
Using the Thom isomorphism, the only non-zero potential contributions to $\tilde{\Omega}^{\rm Spin}_{12}(TF)$ are
$E^2_{11,1} = H_6(F; \mathbb{Z}_2) = \mathbb{Z}_2^4$, and $E^2_{10,2} = H_5(F; \mathbb{Z}_2) = \mathbb{Z}_2$. 

\paragraph{Vanishing of $E^3_{11,1}$} $E^3_{11,1}$ is the cohomology of the sequence
\be
\btkz
H_8(F;\mathbb{Z}) \arrow[r, "d^2_{13,0}"] & H_6(F;\mathbb{Z}_2) \arrow[r,"d^2_{11,1}"] & H_4(F;\mathbb{Z}_2) \;.
\etkz
\ee
We first remark that $\langle h, ({\rm Sq}^2)^\ast ({\rm Sq}^2 h)^\ast \rangle = \langle {\rm Sq}^2 h, ({\rm Sq}^2 h)^\ast \rangle \neq 0$, so $({\rm Sq}^2 h)^\ast$ is not in the kernel of $d^2_{11,1}$ and it is killed by the spectral sequence. Similarly, we compute $Sq^2(w_4) = w_4 w_2$, $Sq^2(w_2^2) = w_3^2$, which means that the kernel of $d^2_{11,1}$ is generated by $(w_2^3)^\ast$.

To check whether $(w_2^3)^\ast$ is in the image of $d^2_{13,0}$, we compute
\be
{\rm Sq}^2(w_2^3) = w_3^2 w_2 + w_2^4 \;.
\ee
As $w_2^4 = \rho_2(p_1^2)$, $\epsilon(w_2^4) = p_1^2 \in {\rm Hom}(H_8(F;\mathbb{Z}), \mathbb{Z}_2)$. Therefore $(d^2_{13,0})^\ast(w_2^3) = p_1^2 + ...$, where the dots denote generators of ${\rm Hom}(H_8(F;\mathbb{Z}), \mathbb{Z}_2)$ independent from $p_1^2$. Therefore $d^2_{13,0}((p_1^2)^\ast) = (w_2^3)^\ast$ and $(w_2^3)^\ast$ is killed as well by the spectral sequence. 

We conclude that $E^3_{11,1} = 0$.

\paragraph{Potential obstruction in $E^3_{10,2}$} $E^3_{10,2}$ is the cohomology of the sequence
\be
\label{EqPotObsSecPageAHSS}
\btkz
H_7(F;\mathbb{Z}_2) \arrow[r, "d^2_{12,1}"] & H_5(F;\mathbb{Z}_2) \arrow[r] & 0 \;.
\etkz
\ee
As ${\rm Sq}^2(w_3w_2) = 0$, we have $(d^2_{12,1})^\ast(w_3w_2) = 0$, so $w_3w_2$ is not in the image of $d^2_{12,1}$ and $(w_3w_2)^\ast$ is not killed on the second page. This generator may be killed by 
\be
\btkz
E^3_{13,0} \arrow[r, "d^3_{13,0}"] & H_5(F;\mathbb{Z}_2) \arrow[r, "d^3_{10,2}"] & E^3_{7,4} = H_2(F;\mathbb{Z}) \;,
\etkz
\ee
but cannot be killed by any of the following differentials. Unfortunately, we don't know what $d^3_{13,0}$ and $d^3_{10,2}$ are.

\subsection{The 12th spin cobordism group of $TF \wedge K(\mathbb{Z},4)$}

We now turn to the second summand of \eqref{EqbordGroup20Struct}. We need first to compute the reduced homology of $TF \wedge K(\mathbb{Z},4)$ with coefficients in $\mathbb{Z}$ and $\mathbb{Z}_2$. Then we can use the Atiyah-Hirzebruch spectral sequence like in the case of $TF$.

The reduced homology of $TF \wedge K(\mathbb{Z},4)$ can be computed with the Künneth short exact sequence:
\be
\btkz
0 \arrow[r] & \bigoplus_{i+j = k} \tilde{H}_i(TF;\mathbb{Z}) \otimes \tilde{H}_{j}(K(\mathbb{Z},4);\mathbb{Z}) \arrow[d] & \\
 & \tilde{H}_{k}(TF \wedge K(\mathbb{Z},4);\mathbb{Z}) \arrow[d] & \\
 & \bigoplus_{i+j = k-1} {\rm Tor}^\mathbb{Z}(\tilde{H}_i(TF;\mathbb{Z}),\tilde{H}_{j}(K(\mathbb{Z},4))) \arrow[r] & 0
\etkz
\ee
The reduced homology of $TF$ is deduced from \eqref{EqHomFIntCoeff} and the Thom isomorphism:
\be
\tilde{H}_\bullet(TF;\mathbb{Z}) = \left( \begin{array}{cccccccccc}
5 & 6 & 7 & 8 & 9 & 10\\
\mathbb{Z} & 0 & \mathbb{Z}_2 & 0 & \mathbb{Z}^2 & \mathbb{Z}_2
\end{array} ... \right) \;.
\ee
The reduced homology of $K(\mathbb{Z},4)$ is given in Appendix C of \cite{Breen2016}:
\be
\tilde{H}_\bullet(K(\mathbb{Z},4);\mathbb{Z}) = \left( \begin{array}{cccccccccc}
4 & 5 & 6 & 7 & 8 & 9\\
\mathbb{Z} & 0 & \mathbb{Z}_2 & 0 & \mathbb{Z} \oplus \mathbb{Z}_3 & 0
\end{array} ... \right) \,.
\ee
(All the reduced homology groups vanish in degrees lower than those indicated.) It turns out that the Tor groups do not contribute in the degrees of interest to us and we find:
\be
\tilde{H}_\bullet(MF \wedge K(\mathbb{Z},4);\mathbb{Z}) = \left( \begin{array}{cccccccccc}
9 & 10 & 11 & 12 & 13\\
\mathbb{Z} & 0 & \mathbb{Z}_2^2 & 0 & \mathbb{Z}^3 \oplus \mathbb{Z}_2 \oplus \mathbb{Z}_3
\end{array} ... \right) \;.
\ee
From this, we deduce the homology with $\mathbb{Z}_2$ coefficients:
\be
\tilde{H}_\bullet(MF \wedge K(\mathbb{Z},4);\mathbb{Z}_2) = \left( \begin{array}{cccccccccc}
9 & 10 & 11 & 12 & 13\\
\mathbb{Z}_2 & 0 & \mathbb{Z}_2^2 & \mathbb{Z}_2^2 & \mathbb{Z}_2^4
\end{array} ... \right) \;.
\ee

We now consider the Atiyah-Hirzebruch spectral sequence for spin bordism
\be
E^2_{p,q} = \tilde{H}_p(TF \wedge K(\mathbb{Z},4), \Omega^{\rm spin}_q({\rm pt.})) \Rightarrow \tilde{\Omega}^{\rm Spin}_{p+q}(TF \wedge K(\mathbb{Z},4)) \;.
\ee
The only potential contribution to $\tilde{\Omega}^{\rm Spin}_{12}(TF \wedge K(\mathbb{Z},4))$ comes from 
\be
E^2_{11,1} = \tilde{H}_{11}(TF \wedge K(\mathbb{Z},4), \mathbb{Z}_2) = \mathbb{Z}_2^2 \;,
\ee
generated by the duals of $u_2 {\rm Sq}^2(g_2)$ and of ${\rm Sq}^2(u_2) g_2$, where $u_2$ is the generator of $H^5(TF;\mathbb{Z}_2)$ and $g_2$ is the generator of $H^4(K(\mathbb{Z},4);\mathbb{Z}_2)$. $E^2_{9,2} = \mathbb{Z}_2$, generated by $u_2 g_2$ and we have
\be
{\rm Sq}^2(u_2 g_2) = {\rm Sq}^2(u_2) g_2 + u_2 {\rm Sq}^2(g_2) \;.
\ee
This means that $d^2_{11,1} = ({\rm Sq}^2)^\ast: E^2_{11,1} \rightarrow E^2_{9,2}$ satisfies
\be
d^2_{11,1}(({\rm Sq}^2(u_2) g_2)^\ast) = d^2_{11,1}((u_2 {\rm Sq}^2(g_2))^\ast) = (u_2 g_2)^\ast \;.
\ee
The kernel of $d^2_{11,1}$ is therefore $({\rm Sq}^2(u_2) g_2)^\ast + (u_2 {\rm Sq}^2(g_2))^\ast$. 

We have
\be
{\rm Sq}^2({\rm Sq}^2(u_2) g_2) = {\rm Sq}^2(u_2 {\rm Sq}^2(g_2)) = {\rm Sq}^2(u_2) {\rm Sq}^2(g_2) \;.
\ee
If we can show that $\epsilon({\rm Sq}^2(u_2) {\rm Sq}^2(g_2)) \neq 0$, we will have shown that $({\rm Sq}^2(u_2) g_2)^\ast + (u_2 {\rm Sq}^2(g_2))^\ast$ is in the image of $d^2_{13,0}$, and therefore that $E^2_{11,1}$ is killed by the spectral sequence.

The universal coefficient theorem computing $\mathbb{Z}_2$-valued cohomology together with the fact that $H_6(MF;\mathbb{Z}) = 0$ show that $\epsilon({\rm Sq}^2(u_2))$ is the non-trivial element of ${\rm Hom}(H_7(MF;\mathbb{Z}),\mathbb{Z}_2)$. Similarly, as $H_5(K(\mathbb{Z},4);\mathbb{Z}) = 0$, $\epsilon({\rm Sq}^2(g_2))$ is the nontrivial element of ${\rm Hom}(H_6(K(\mathbb{Z},4);\mathbb{Z}),\mathbb{Z}_2)$. This implies that $\epsilon({\rm Sq}^2(u_2) {\rm Sq}^2(g_2)) \neq 0$.

We conclude that $\Omega^{\rm Spin}_{12}(TF \wedge K(\mathbb{Z},4)) = 0$, and therefore that
\be
\Omega^{M5}_{12} = \mathbb{Z}_2 \mbox{ or } 0\;,
\ee
where the uncertainty comes from the cokernel of $d^2_{12,1}$ in the Atiyah-Hirzebruch spectral sequence for $\Omega^{\rm Spin}_{12}(TF)$, see \eqref{EqPotObsSecPageAHSS}.

The computations above show that if $\Omega^{M5}_{12}$ is non-trivial, its only generator is the dual of $w_2(\mathscr{N})w_3(\mathscr{N})$. Therefore any $(2,0)$-manifold such that $w_2(\mathscr{N})w_3(\mathscr{N}) = 0$ bounds, proving the theorem.

{
\small

\providecommand{\href}[2]{#2}\begingroup\raggedright\endgroup

}

\end{document}